\documentclass[
reprint,
superscriptaddress,
 amsmath,amssymb,
 prl,
]{revtex4-2}

\usepackage{graphicx}
\usepackage{dcolumn}
\usepackage{bm}
\usepackage{hyperref}
\hypersetup{
    colorlinks=true,
    allcolors=blue
}
\usepackage[mathlines]{lineno}
\usepackage{siunitx}
\usepackage{csquotes}
\usepackage{braket}
\usepackage{wasysym} 
 \usepackage{color}

\newcommand{\beginsupplement}{%
        \setcounter{table}{0}
        \renewcommand{\thetable}{S\arabic{table}}%
        \setcounter{figure}{0}
        \renewcommand{\thefigure}{S\arabic{figure}}%
        \setcounter{equation}{0}
     }

\begin{document}

\title{Coherent control of a symmetry-engineered multi-qubit dark state in waveguide quantum electrodynamics}



\author{Maximilian Zanner}
\email{maximilian.zanner@uibk.ac.at}
\affiliation{Institute for Experimental Physics, University of Innsbruck, 6020 Innsbruck, Austria}
\affiliation{Institute for Quantum Optics and Quantum Information, Austrian Academy of Sciences, 6020 Innsbruck, Austria}

\author{Tuure Orell}
\affiliation{Nano and Molecular Systems Research Unit, University of Oulu, 90014 Oulu, Finland}

\author{Christian M. F. Schneider}
\affiliation{Institute for Experimental Physics, University of Innsbruck, 6020 Innsbruck, Austria}
\affiliation{Institute for Quantum Optics and Quantum Information, Austrian Academy of Sciences, 6020 Innsbruck, Austria}

\author{Romain Albert}
\affiliation{Institute for Experimental Physics, University of Innsbruck, 6020 Innsbruck, Austria}
\affiliation{Institute for Quantum Optics and Quantum Information, Austrian Academy of Sciences, 6020 Innsbruck, Austria}

\author{Stefan Oleschko}
\affiliation{Institute for Experimental Physics, University of Innsbruck, 6020 Innsbruck, Austria}
\affiliation{Institute for Quantum Optics and Quantum Information, Austrian Academy of Sciences, 6020 Innsbruck, Austria}

\author{Mathieu L. Juan}
\affiliation{Institut Quantique/Département de physique, Université de Sherbrooke, Sherbrooke, Québec J1K 2R1, Canada}

\author{Matti Silveri}
\affiliation{Nano and Molecular Systems Research Unit, University of Oulu, 90014 Oulu, Finland}

\author{Gerhard Kirchmair}
\affiliation{Institute for Experimental Physics, University of Innsbruck, 6020 Innsbruck, Austria}
\affiliation{Institute for Quantum Optics and Quantum Information, Austrian Academy of Sciences, 6020 Innsbruck, Austria}

\date{\today}

\begin{abstract}
Quantum information is typically encoded in the state of a qubit that is decoupled from the environment. In contrast, waveguide quantum electrodynamics studies qubits coupled to a mode continuum, exposing them to a loss channel and causing quantum information to be lost before coherent operations can be performed. Here we restore coherence by realizing a dark state that exploits symmetry properties and interactions between four qubits. Dark states decouple from the waveguide and are thus a valuable resource for quantum information but also come with a challenge: they cannot be controlled by the waveguide drive. We overcome this problem by designing a drive that utilizes the symmetry properties of the collective state manifold allowing us to selectively drive both bright and dark states. The decay time of the dark state exceeds that of the waveguide-limited single qubit by more than two orders of magnitude. Spectroscopy on the second excitation manifold provides further insight into the level structure of the hybridized system. Our experiment paves the way for implementations of quantum many-body physics in waveguides and the realization of quantum information protocols using decoherence-free subspaces.
\end{abstract}
\maketitle


\begin{figure*}[ht]
    \centering
    \includegraphics[width=\linewidth]{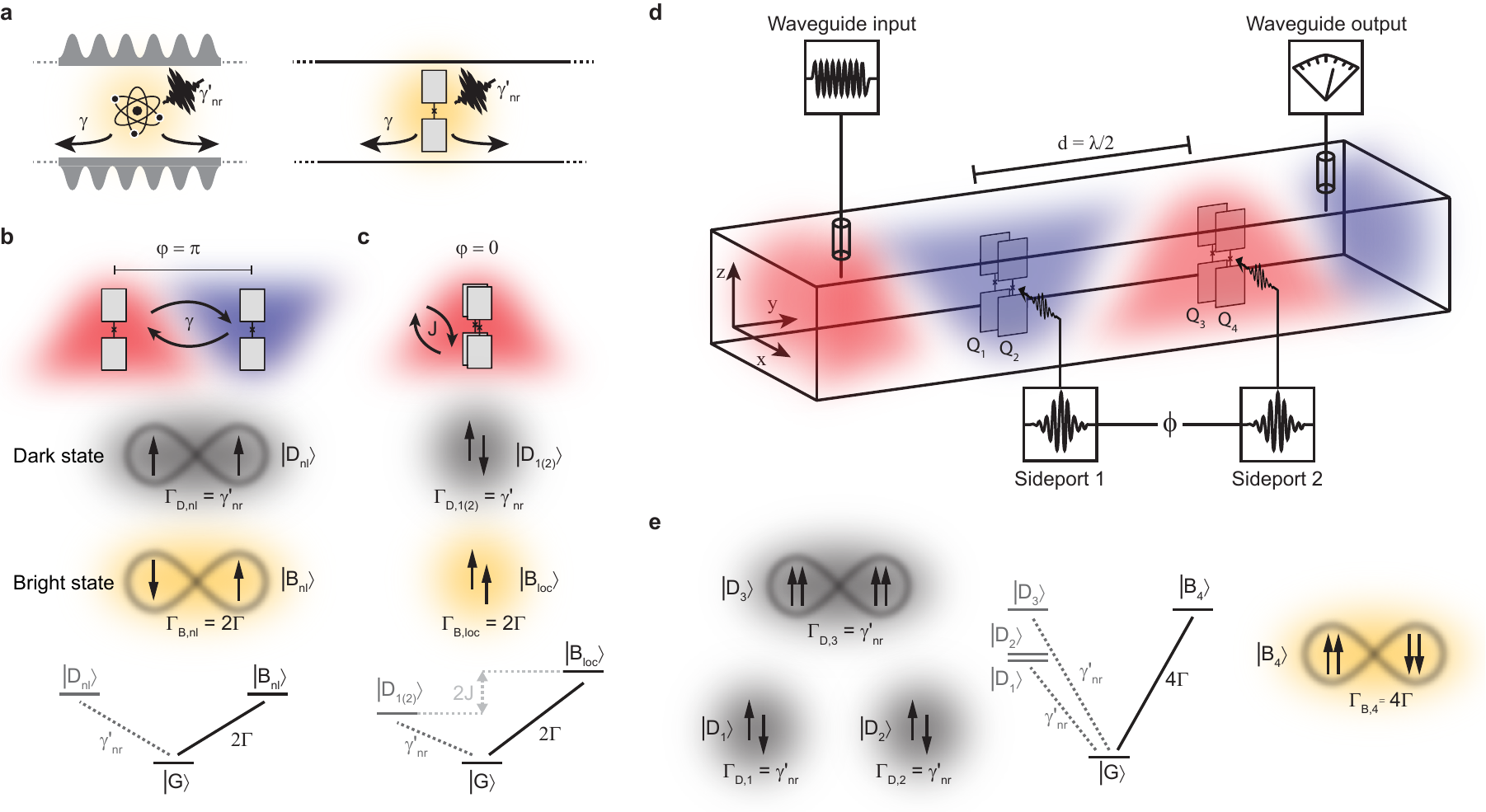}
    \caption{\textbf{Symmetry of collective states in waveguide quantum electrodynamics.}
    \textbf{a} Left: A natural atom in a photonic crystal waveguide. Right: A transmon qubit, acting as an artificial atom is coupled to a mode continuum in superconducting waveguide quantum electrodynamics. In the strong coupling limit the decay into the waveguide $\gamma$ dominates over non-radiative losses $\gamma_{\rm nr}'$.
    \textbf{b} Depending on the symmetry of the electromagnetic environment, multi-qubit collective states acquire super- or subradiant behaviour depending on their respective symmetry. The symmetries of the collective states are indicated by arrows representing in-phase and out-of-phase oscillating transition dipole moments. For two qubits separated by $d = \lambda/2$, the dark state is a symmetric superposition $\ket{D_{\rm nl}} = \left( \ket{eg} + \ket{ge} \right)/\sqrt{2}$ and the bright state $\ket{B_{\rm nl}} = \left( \ket{eg} - \ket{ge} \right)/\sqrt{2}$ is an antisymmetric superposition. The phase and amplitude of the field in the waveguide is sketched with the red and blue shaded regions.
    \textbf{c} Two directly coupled qubits with no separation along the propagation direction, i.e. $\varphi = 0$, form an antisymmetric dark state $\ket{D_{1(2)}} = \left( \ket{eg} - \ket{ge} \right)/\sqrt{2}$ and a symmetric bright state $\ket{B_{\rm loc}} = \left( \ket{eg} + \ket{ge} \right)/\sqrt{2}$. The dark state indices denote, that there are two pairs that create two local dark states. In contrast to the waveguide mediated coupling, the degeneracy of the energy of the bright and dark state is lifted by twice the coherent exchange coupling rate $2 J$.
    \textbf{d} Schematic illustration of the full setup. Two pairs of transmon qubits are separated by an effective distance $d = \lambda/2$. All states can be addressed by the ports in the sidewall of the waveguide (Sideports 1 and 2).
    \textbf{e} Two-transmon bright states of local pairs interact via the waveguide to form  the four-qubit dark state $\ket{D_3}$ and bright state $\ket{B_4}$ with decay rate $4 \Gamma$. Pairwise dark states $\ket{D_1},\ket{D2}$ are localized at the sites and do not interact with the waveguide or the other pair.}
    \label{fig:Fig1}
\end{figure*}

Waveguide quantum electrodynamics has become a popular platform to study light-matter interaction by coupling atoms to a one dimensional continuum of modes~\cite{roy_colloquium_2017, masson_atomic-waveguide_2020, goban_atomlight_2014, corzo_large_2016, lodahl_interfacing_2015,kannan_waveguide_2020}. Superconducting qubits are in many ways ideal emitters for waveguide quantum electrodynamics. They can act as artificial atoms and are highly engineerable such that they can be strongly coupled to the one-dimensional mode continuum of a waveguide, where the coupling rate $\gamma$ dominates over all other decoherence channels $\gamma_{\rm nr}'$. These properties lead to the observation of a  broad range  of  phenomena  such  as  the Mollow triplet \cite{astafiev_resonance_2010-2,baur_measurement_2009}, ultra strong coupling \cite{forn-diaz_ultrastrong_2017}, generation  of non-classical photonic states~\cite{hoi_generation_2012, kannan_generating_2020-1}, qubit-photon bound states \cite{sundaresan_interacting_2019}, chiral \cite{lodahl_chiral_2017-1} and topological physics \cite{kim_quantum_2021} as well as collective effects \cite{loo_photon-mediated_2013}.

Collective states appear naturally in waveguide quantum electrodynamics and are a result of waveguide-mediated interactions and interference effects between individual emitters~\cite{lalumiere_input-output_2013-2, kockum_decoherence-free_2018}. Each collective state gives rise to an effective dipole moment, where the symmetry of the electric field with respect to the waveguide mode determines whether the collective state obtains a sub- or  superradiant  decay  rate, i.e whether it becomes a dark or bright state. Dark states effectively form decoherence-free subspaces~\cite{monz_realization_2009,kwiat_experimental_2000,kielpinski_decoherence-free_2001} that introduce the possibility to realize a quantum computation and simulation platform within an open system ~\cite{lidar_decoherence-free_1998, paulisch_universal_2016-2}. Collective bright and dark states have already been realized in various superconducting qubit systems~\cite{loo_photon-mediated_2013, mlynek_observation_2014-3, rosario_hamann_nonreciprocity_2018, mirhosseini_cavity_2019-1} but so far coherent control of a multi-qubit dark state has not been achieved. The difficulty arises from the main property of the dark state - it decouples from the environment. 

In this work we realize a collective dark state qubit by coupling four superconducting transmon qubits to the mode continuum of a rectangular waveguide. The collective dark state can be coherently controlled with two physically separate drive ports, which allows us to adjust the symmetry of the drive and thus solves the problem of driving a state that is decoupled from the waveguide. By carefully tuning the qubits to the decoherence-free subspace, we demonstrate long coherence time along with coherent control in an open quantum system. In particular, we achieve an effective protection from the waveguide, leading to a decrease of the relaxation rate by a factor of 320 compared to the single qubit coupling rate, or a factor of 1300 compared to the collective bright state. We utilize the bright state to read out the ground state population in this resonator-free setup. Moreover, we perform a pulsed spectroscopy on the second excitation manifold to characterize other collective states and use superradiant transitions to reset the dark state qubit.

\begin{figure*}[ht]
    \centering
    \includegraphics[width=1\linewidth]{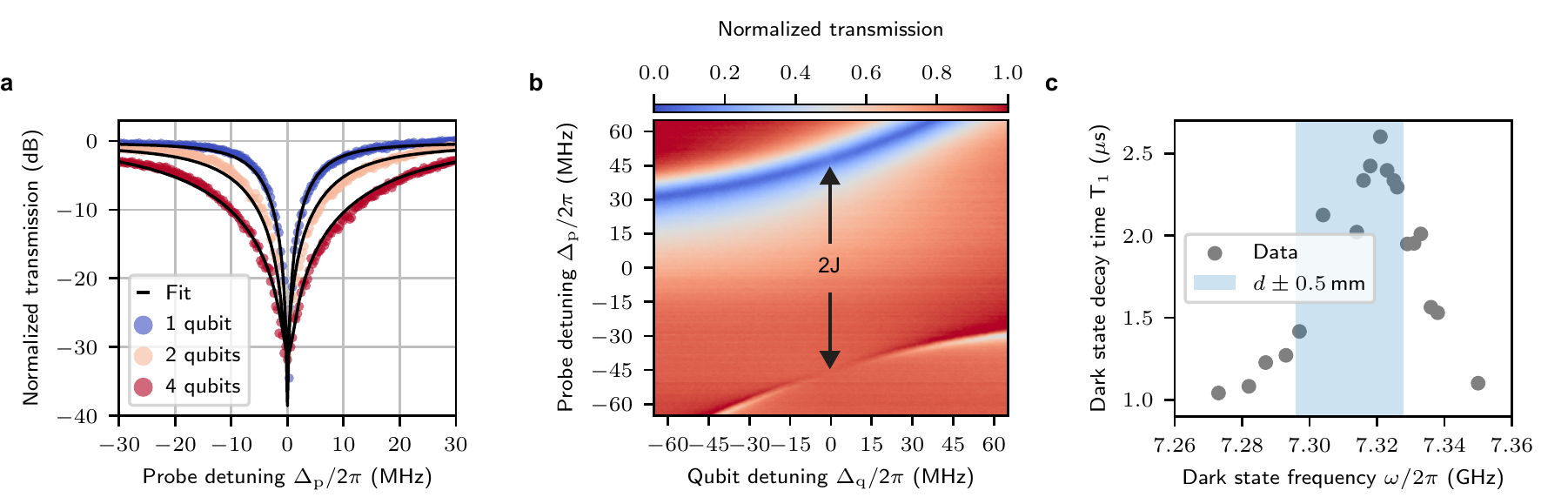}
    \caption{\textbf{Measuring dark and bright states.}
    \textbf{a} Waveguide transmission as a function of probe frequency around a resonance where the dip corresponds to a single qubit or the common bright state of the hybridized two and four transmon systems. The single qubit linewidth $\Gamma/2\pi = \SI{14.9}{\mega\hertz}$ increases to $\Gamma_{\rm B}/2\pi = \SI{30.2}{\mega\hertz}$ for the two-qubit (local or non-local) bright state and to $\Gamma_{\rm B,4}/2\pi = \SI{60.9}{\mega\hertz}$ for the four qubit state  corresponding to superradiant transitions.
    \textbf{b} Waveguide transmission as a function of qubit and probe detuning. Two transmon qubits $Q_3$ and $Q_4$ (equivalent for $Q_1$ and $Q_2$, not shown) are tuned in and out of resonance while measuring transmission through the waveguide for different frequencies. From the avoided crossing we extract direct capacitive coupling strengths $J_{12} = 2\pi\times\SI{43}{\mega\hertz}$ and $J_{34} = 2\pi\times\SI{47}{\mega\hertz}$. The upper branch obtains twice the linewidth $\Gamma_{\rm B, loc}=2\Gamma$ whereas the lower branch disappears from the transmission, effectively decoupling the transition from the waveguide and reducing the dark state decay rate to non-radiative losses $\Gamma_{\rm D,1(2)}=\gamma_{\rm nr}'$.
    \textbf{c} Dark state lifetime for a $\lambda/2$ separated transmon pair as a function of dark state frequency. The frequency for maximal correlated decay $\omega_{\pi}/2\pi = \SI{7.321}{\giga\hertz}$ is calibrated by measuring dark state decay times around the analytical value for a physical separation of $d_y = \SI[separate-uncertainty]{46\pm0.5}{\milli\meter}$ (blue region). Correlated decay depends on the physical separation and the wavelength. When the frequency fulfills $\lambda/2 =2\pi \mathrm{v}/\omega_{j}$, the dark state symmetry is optimal, hence we measure the longest decay time T$_1$.
    }
    \label{fig:Fig2}
\end{figure*}
The decoherence rate of a qubit with ground state $\ket{g}$ and excited state $\ket{e}$ is given by its linewidth. As illustrated in Fig.~\hyperref[fig:Fig1]{1a}, the qubit decoherence rate $\Gamma = (\gamma + \gamma_{\rm nr})/2 + \gamma_{\phi} $ is a sum of radiative decay $\gamma$ into the waveguide modes, non-radiative energy loss $\gamma_{\rm nr}$, as well as pure dephasing $\gamma_{\phi}$. When the qubit is coupled to a waveguide, the linewidth can be extracted in scattering experiments by measuring the waveguide transmission or reflection, allowing to obtain $\gamma$ and the non-radiative decoherence rate $\gamma_{nr}' = \gamma_{\rm nr}/2 + \gamma_{\phi}$. The dimensionless coupling strength to the waveguide $g_j=\sqrt{\gamma_{j}/(2\pi\omega_{j})}$~\cite{lalumiere_input-output_2013-2} is then given by the radiative decay, normalized by the resonance frequency $\omega_{j}$ of qubit $j$.

The device sketched in Fig.~\hyperref[fig:Fig1]{1d} consists of four frequency-tunable transmon qubits~\cite{koch_charge-insensitive_2007-2} acting as artificial atoms. Transmons $Q_1$ and $Q_2$ are located on the left, closer to the input side of the waveguide. Transmons $Q_3$ and $Q_4$ are located on the right, closer to the output of the waveguide, such that the physical separation between the pairs is $d_y = \SI[separate-uncertainty]{46\pm0.5}{\milli\meter}$. Within the pairs, the transmons are separated by $d_x = \SI{1}{\milli\meter}$ which gives rise to a capacitive coupling. The fundamental waveguide mode $\mathrm{TE_{01}}$ has a cutoff-frequency of $\omega_{c}/2\pi=\SI{6.55}{\giga\hertz}$~\cite{pozar_microwave_2012} and a polarization of the electrical field that is parallel to the dipole moment of the transmons, such that they efficiently couple to the waveguide.

We study the case where two transmons interact through the waveguide at a separation $d_y=\lambda/2$ in Fig.~\hyperref[fig:Fig1]{1b}. The signal propagating between the transmons acquires a phase $\varphi = 2\pi d_y / \lambda$ depending on the wavelength $\lambda=2\pi \mathrm{v} /\omega_{j}$ and separation $d_y$, where $\mathrm{v}$ is the speed of light in the waveguide (see Supplementary Material). Analytically, a phase difference of $\varphi=\pi$ for our setup corresponds to an emission frequency $\omega_{\pi}/2\pi = \SI[separate-uncertainty]{7.312 \pm 0.016}{\giga\hertz}$. There, correlated decay into waveguide modes $\gamma_{j,k} = 2\pi g_j g_k \omega_{j} \cos(\varphi)$ is maximized~\cite{lalumiere_input-output_2013-2} and coherent waveguide-mediated interaction is absent, due to the counter-periodic behavior of $\widetilde{J}_{j,k} = \pi g_j g_k \omega_{j} \sin(\varphi)$. The dissipative interaction leads to symmetric and antisymmetric states under qubit exchange, i.e. the dark state $\ket{D_{\rm nl}} = \left( \ket{ge} + \ket{eg} \right)/\sqrt{2}$ and bright state $\ket{B_{\rm nl}} = \left( \ket{ge} - \ket{eg} \right)/\sqrt{2}$. For a distance of $\lambda/2$, the phase relation of the electromagnetic field in the waveguide is antisymmetric ($\varphi=\pi$), thus we can only excite the antisymmetric bright state. The dark state symmetry is opposite to the field symmetry of the waveguide, eliminating the coupling to the drive field and decay into waveguide modes.

Two nearby transmons are directly coupled through the capacitance between the metallic pads of their antennae. Unlike interactions mediated by the waveguide, the capacitive coupling for transmons in this configuration has an effective $1/r^3$-dependence~\cite{dalmonte_realizing_2015-5}, leading to short range coupling. On resonance, an excitation can swap coherently between the local transmons, resulting in new eigenstates, in particular a symmetric state $\ket{B_{\rm loc}} = \left( \ket{ge} + \ket{eg} \right)/\sqrt{2}$ and an antisymmetric state $\ket{D_{1(2)}} = \left( \ket{ge} - \ket{eg} \right)/\sqrt{2}$, illustrated in Fig.~\hyperref[fig:Fig1]{1c}. The capacitively coupled transmons are located at the same position with respect to the propagating field and symmetrically around the center of the waveguide. Therefore, the phase of the electrical field is the same for both transmons $\varphi = 0$ and the drive along the waveguide can only access the symmetric state, in contrast to the scenario where the qubits are separated by $\lambda/2$.

The individual qubit and collective bright state decay rates are extracted from transmission measurements, using a circle-fit routine~\cite{probst_efficient_2015-3} on the complex-valued scattering parameters. In Fig.~\hyperref[fig:Fig2]{2a}, we show the magnitude of the normalized transmission for a single transmon with emission frequency $\omega_{\pi}$, as well as for two and four transmons.

The capacitively coupled transmon pairs obtain direct coupling strengths $J_{12} =2\pi\times \SI{43}{\mega\hertz}$ and $J_{34} = 2\pi\times \SI{47}{\mega\hertz}$ which can be extracted from an avoided crossing, shown for $Q_3$ and $Q_4$ in Fig.~\hyperref[fig:Fig2]{2b}. The difference in coupling strengths is a result of imperfections in the alignment. The coherent exchange interaction lifts the degeneracy of $\ket{B_{\rm loc}}$ and $\ket{D_{1(2)}}$ and allows us to observe the decoupling of the dark state when we tune the qubits into resonance.

The long-lived nature of dark states comes from the fact that they decouple from the mode environment which means that resonant driving via the waveguide is not possible. In order to achieve control of the dark states we introduce two weakly coupled sideports, sketched in Fig.~\hyperref[fig:Fig1]{1d}. They provide an amplitude gradient over the local pairs to access dark states $\ket{D_{\rm 1(2)}}$, but also the possibility to independently adjust the phase $\phi$, which allows us to apply a symmetric drive and access the non-local dark state $\ket{D_{\rm nl}}$. To ensure the locality of the drive, both ports are engineered such that the electrical field is perpendicular to the $\rm TE_{10}$ mode of the waveguide. 

We can measure the ground state population by employing a scheme similar to the electron shelving method used for quantum non-demolition state detection in trapped ion quantum computing \cite{leibfried_quantum_2003}. If the collective system is in the ground state $\ket{G}$ we can coherently scatter photons between the ground state $\ket{G}$ and superradiant state $\ket{B_{\rm nl}}$, which reduces the transmission through the waveguide, as can be seen in Fig.~\hyperref[fig:Fig2]{2a}. On the other hand, if the dark state $\ket{D_{\rm nl}}$ is populated, the microwave signal is not scattered, resulting in unit transmission. By selectively exciting the dark state using microwave signals applied through the sideports with $\phi=0$ we can experimentally search for the longest dark state relaxation time around the decoherence-free frequency and therefore calibrate the frequency corresponding to $d=\lambda/2$ at $\omega_{\pi}/2\pi = \SI{7.321}{\giga\hertz}$, shown in Fig.~\hyperref[fig:Fig2]{2c}. 

\begin{figure*}[th]
    \centering
    \includegraphics[width=1\linewidth]{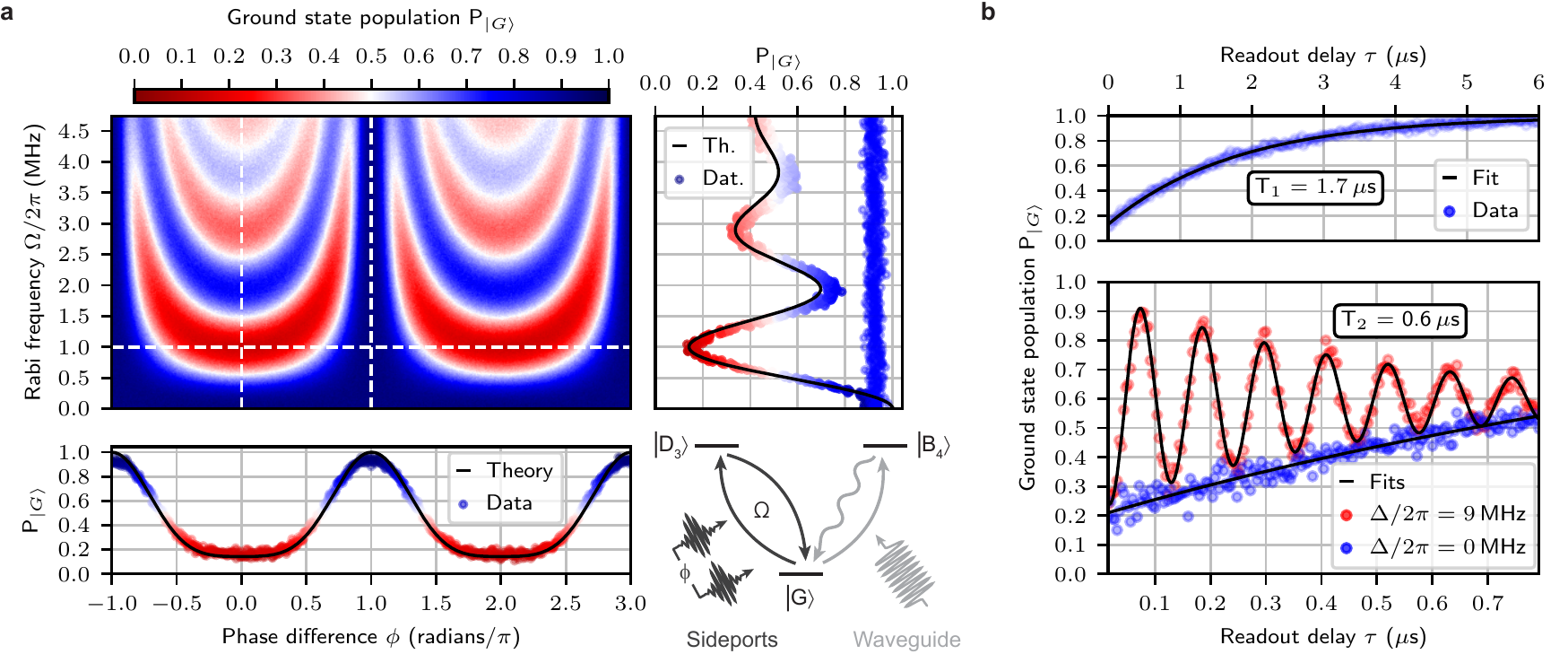}
    \caption{\textbf{Coherent control of the decoherence-free subspace.}
    \textbf{a} We apply a gaussian shaped pulse of length $\rm t_{a} = \SI{240}{\nano\second}$ (full width $\sigma_{a} = \rm t_{a}/3$) to observe Rabi oscillation between the ground state $\ket{G}$ and the non-local four qubit dark state $\ket{D_3}$ as a function of the Rabi frequency $\Omega$, varied by the drive amplitude and the sideport phase difference $\phi$. By applying the pulse through the sideports we can set the phase $\phi$ independently. The ground state population is read out by sending a \SI{5}{\micro\second} long rectangular pulse through the waveguide, resonant with the transition between states $\ket{G}$ and $\ket{B_4}$. The right panel shows a vertical linecut at the white dashed lines of the colormap for phase-difference $\phi =0$ and $\phi = 1$. The lower panel shows a horizontal linecut for a Rabi-frequency of $\Omega/2\pi = \SI{1}{\mega \hertz}$. For the theory curve, we simulate the Hamiltonians Eqs.~\eqref{eq:transmon_hamilton} to \eqref{eq:driving_hamiltonian} and master equations in the Supplementary Material with the parameters specified in Supplementary Tab.~\ref{tab:transmonpars}.
    \textbf{b} A symmetric excitation pulse with Rabi frequency $\Omega/2\pi = \SI{1}{\mega\hertz}$ and relative phase difference $\phi =0$ between both sideports is used to populate the collective dark state $\ket{D_3}$. After a variable delay time, the ground state population is read out to find an average relaxation time $T_1 = \SI[separate-uncertainty]{1.71\pm0.06}{\micro\second}$. In a Ramsey experiment we find an average coherence time of $T_2 = \SI[separate-uncertainty]{0.58\pm0.06}{\micro\second}$. On resonance $\Delta/2\pi=0$ we observe an exponential decay and for a detuned pulse $\Delta/2\pi= \SI{9}{MHz}$ we induce oscillations with frequencies corresponding to the detuning between drive frequency and transition frequency between the states $\ket{G}$ and $\ket{D_3}$.}
    \label{fig:Fig3}
\end{figure*}

Next, we tune all four transmons into resonance such that the bright transitions of the capacitively coupled pairs match the decoherence-free frequency $\omega_{\pi}$. Both local two-qubit bright states interact via the waveguide and create the collective four qubit states $\ket{B_4}$ and $\ket{D_3}$, whereas the local two-qubit dark states $\ket{D_1}$ and $\ket{D_2}$ cannot interact via the waveguide. These four states span the first excitation manifold, depicted in Fig.~\hyperref[fig:Fig1]{1e}. In Fig.~\hyperref[fig:Fig2]{2a} we extract the linewidth $\Gamma_{\rm B,4}/2\pi = \SI{60.9}{\mega\hertz}$ resulting from constructive interference of all transmons $\Gamma_{\rm B,4} = \sum_j \Gamma_{j}$. The symmetric superposition of the two-qubit bright states interfere destructively and isolate state $\ket{D_3}$ from the waveguide.

In the following we want to investigate time-resolved dynamics when driving the system through the sideports. Rabi-oscillations between $\ket{G}$ and $\ket{D_3}$ can be observed in Fig.~\hyperref[fig:Fig3]{3a} when the amplitude of the drive field $\Omega$ is increased and the phase difference between the sideports matches that of the non-local dark state: $\phi = 2n\pi$. For an antisymmetric drive with odd integer multiple $\phi = (2n-1) \pi$, we only drive the bright state $\ket{B_4}$ which decays very rapidly to the ground state with the rate $\Gamma_{\rm B,4}$. For phases that are neither fully symmetric nor antisymmetric we drive both states simultaneously, where the respective drive strength depends on the phase. Again, we employ the electron shelving readout scheme as for the two-qubit case, now using transition $\ket{G}$ to $\ket{B_4}$ to scatter waveguide photons. With a calibrated $\pi$ and $\pi/2$-pulse, we can investigate the coherence properties of the dark state. For the collective dark state we measure an average relaxation time T$_1 = \SI[separate-uncertainty]{1.71\pm0.06}{\micro\second}$ and coherence time T$_2 = \SI[separate-uncertainty]{0.58\pm0.06}{\micro\second}$, shown in Fig.~\hyperref[fig:Fig3]{3b}. In this system, dephasing and frequency fluctuations of the individual qubits cause imperfections of the dark state symmetry. This results in a finite decay rate into the waveguide, thus $\rm T_1$ depends on $\gamma_\phi$ (see Supplementary Material).

To simulate the collective dynamics shown by the black lines in Fig.~\hyperref[fig:Fig3]{3a}, we model the transmons and their direct couplings with the Hamiltonian~\cite{Roushan17, Ma2019, hacohen-gourgy_cooling_2015}
\begin{equation}\label{eq:transmon_hamilton}
    \begin{aligned}
        \hat H_{\rm T}/\hbar &= \sum_{j=1}^4\left[
        \omega_j\hat n_j - \frac{U_j}{2}\hat n_j(\hat n_j-1)\right]\\
        &+J_{\rm 12}\left(\hat a_1^\dag\hat a_2 +\rm{h.c.}\right)
        +J_{\rm 34}\left(\hat a_3^\dag\hat a_4 +\rm{h.c.}\right),
    \end{aligned}
\end{equation}
where $\omega_j$ are the fundamental resonance frequencies and $U_j$ the anharmonicities of the individual transmons. Operators $\hat a_j$ and $\hat a_j^\dag$ are the annihilation and creation operators of the $j$th transmon and $\hat n_j=\hat a_j^\dag\hat a_j$ is the corresponding number operator. In the presence of the waveguide radiation field, the dynamics is governed by a master equation~\cite{lalumiere_input-output_2013-2, mirhosseini_cavity_2019-1} taking into account the coherent exchange interaction $\widetilde{J}_{j,k}$ and the correlated decay $\gamma_{j,k}$ between the transmons at sites $j$ and $k$ (see Supplementary Material). The properties of the system are then described by the effective non-Hermitian Hamiltonian,
\begin{equation}\label{eq:effective_hamiltonian}
    \begin{aligned}
        \hat H_{\rm eff}/\hbar &= \hat H_{\rm T}/\hbar
        +\sum_{jk}\left(\widetilde{J}_{j,k} -\frac{i\gamma_{j,k}}{2}\right)
        \hat a_k^\dag\hat a_j\\
        &-\frac{i}{2}\sum_j\gamma_{\rm nr}\hat a_j^\dag\hat a_j.
    \end{aligned}
\end{equation}
with the parameter $\gamma_{\rm nr}$ describing the non-radiative dissipation of individual transmons.
The eigenvalues of the effective Hamiltonian are complex valued $\lambda_\alpha = E_\alpha-i\frac{\Gamma_\alpha}{2}$, where the real part gives the energy $E_\alpha$ and the imaginary part the total decay rate $\Gamma_{\alpha}$ of state $\ket{\alpha}$. By analyzing the real and imaginary parts, we can identify the dark and bright states of the collective system. The effective Hamiltonian commutes with the total occupation operator, thus the eigenvalues form manifolds with integer number of quanta. The eigenstates of the first excitation manifold are either symmetric or antisymmetric with respect to the exchange of transmon pairs. The second excitation manifold also comprises states that cannot be assigned to a pair-exchange symmetry, but instead they are symmetric or antisymmetric with respect to the exchange of transmons within the pairs (see Supplementary Material for details).

In the frame rotating with the drive frequency $\omega$, the simplified driving Hamiltonian reads
\begin{equation}\label{eq:driving_hamiltonian}
    \hat H_{\rm d}(t)/\hbar = \frac{\Omega(t)}{2}\left[e^{i\phi}
    \left(\hat a_1+\hat a_2\right) 
    + \hat a_3+\hat a_4 + \rm{h.c.}\right].
\end{equation}
The phase alters the symmetry with respect to the exchange of the pairs, but the drive is always symmetric with respect to the exchange of transmons within the pairs. By modifying the phase one can thus couple to different states in the neighbouring manifolds. Most importantly, one can show that the drive strength from ground to bright state, as well as ground to dark state depends on the phase~$\phi$
\begin{align}
    \braket{D_3|\hat H_{\rm d}|G} &= \frac{\hbar\Omega}{2}\left(1+e^{i\phi}\right),\\
    \braket{B_4|\hat H_{\rm d}|G} &= \frac{\hbar\Omega}{2}\left(1-e^{i\phi}\right).
\end{align}
With $\phi=0$, one can therefore only drive the dark state $\ket{D_3}$ and similarly with $\phi=\pi$ only the bright state $\ket{B_4}$.

\begin{figure*}[ht]
    \centering
    \includegraphics[width=1\linewidth]{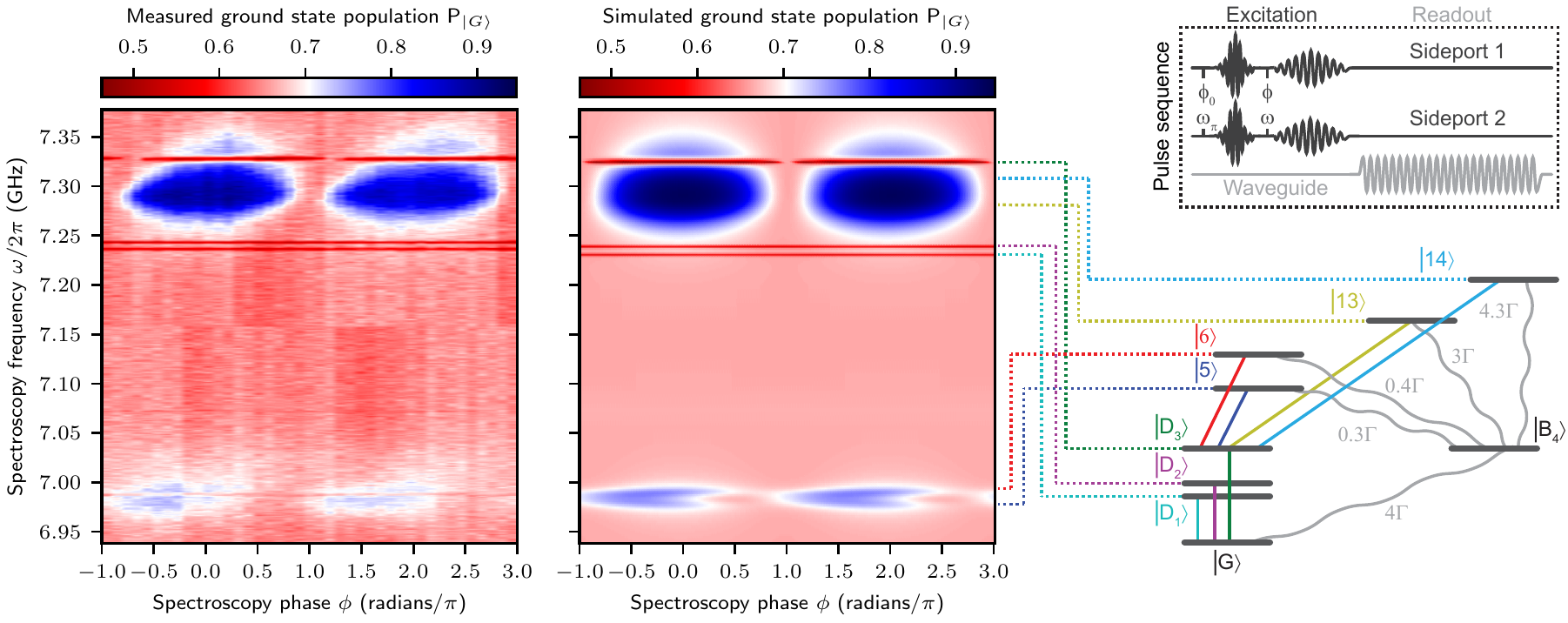}
    \caption{\textbf{Phase-sensitive spectroscopy of the second excitation manifold.}
    In the upper right panel we show the pulse sequence where we apply a gaussian shaped pulse of length $\rm t_{a} = \SI{240}{\nano\second}$ (full width $\sigma_{a} = \rm t_{a}/3$) that brings most of the population to the dark state $\ket{D_3}$. Here, the relative phase on the sideports is set to $\phi_0=0$ and the frequency is set to the decoherence-free frequency $\omega_{\pi}$. Afterwards, we concatenate a gaussian spectroscopy pulse of length $\rm t_{b} = \SI{1.2}{\micro\second}$ (full width $\sigma_{b} = \rm t_{b}/3$) with variable frequency $\omega/2\pi$ and phase $\phi$. We record the waveguide transmission with a rectangular readout pulse of length $t_c=\SI{5}{\micro \second}$ to obtain the measured (left) and simulated (right) ground state population. The states in the second excitation manifold have a finite decay rate to the bright state $\ket{B_4}$, which again decays to the ground state $\ket{G}$. Therefore we measure a high ground state population when the spectroscopy pulse is resonant with a transition that can be driven from $\ket{D_3}$. The ability to drive collective states depends on the spectroscopy phase whereas local states can be driven with any phase. The parameters for the simulation are given in the Supplementary Tab.~\ref{tab:transmonpars}. In order observe the local dark states in the simulation, we have included an amplitude gradient of the local drives, such that the power on transmons $Q_2$ and $Q_4$ is three quarters of that on $Q_1$ and $Q_3$.}
    \label{fig:Fig4}
\end{figure*}

To fully utilize the decoherence-free subspace of the ground state $\ket{G}$ and the dark state $\ket{D_3}$, we need to understand how well it is isolated from the higher excitation manifolds. Here it is essential to notice that already a single transmon is a multilevel system with anharmonicity $U$. We can explore the second excitation manifold of the collective four-transmon system in Fig.~\hyperref[fig:Fig4]{4} by concatenating a spectroscopy pulse after populating the dark state $\ket{D_3}$, using a $\pi$-pulse with a phase difference $\phi=0$. For the spectroscopy we change the frequency and the relative phase to unveil the symmetry and energy of the states in the second excitation manifold. When the spectroscopy pulse is resonant with a transition to a state of the second excitation manifold, e.g. $\ket{5}$, $\ket{6}$, $\ket{13}$ or $\ket{14}$ (see Supplementary Material) the system is reset to the ground state due to the rapid decay of these states via the bright state $\ket{B_4}$. In Fig.~\hyperref[fig:Fig4]{4}, the collectiveness of these states is apparent in the phase dependence of the measured ground state population (left panel), which is consistent with the simulation (right panel) of the model Hamiltonian Eq.~\eqref{eq:effective_hamiltonian}. During the time of the spectroscopy pulse, which is on the same order as the lifetime of $\ket{D_3}$, a part of the population decays to the ground state. As a consequence, the phase sensitive transition between states $\ket{G}$ and $\ket{D_3}$ is visible as well. In addition, the dark states $\ket{D_1}$ and $\ket{D_2}$ are also visible in the spectroscopy, as there is an amplitude gradient across a pair. This asymmetry produces an additional driving term that is always antisymmetric with respect to the exchange of transmons within the pair. These states do not show a phase dependence as they are only coupled to one drive port. 

The spectroscopy in Fig.~\hyperref[fig:Fig4]{4} shows that the linewidth of the transitions $\ket{D_3}$ to $\ket{13}$ and $\ket{14}$ is larger than the detuning of their resonance frequency with respect to the transition $\ket{G}$ to $\ket{D_3}$. Consequently, the drive populates these states in the second excitation manifold and we observe damped Rabi oscillations in Fig.~\hyperref[fig:Fig3]{3a} when we increase the amplitude of the excitation pulse. Remarkably, this leakage effect can be reduced dramatically by increasing the coupling to the waveguide so much that the unwanted excitation to this state can be adiabatically eliminated (Supplementary Material). The strong coupling of the second excitation manifold to the waveguide can be utilized to cool and reset the collective dark state qubit and deterministically generate itinerant photons. Notice that, ideally, increasing the waveguide coupling does not affect the coherence and lifetime of the dark state, only the states outside the decoherence-free subspace decay faster. In contrast to conventional solid-state qubits, a symmetry-engineered multi-qubit system makes it possible to control the decay properties of the leakage states independently of the computational states.

In conclusion, we engineered a collective four-qubit dark state in a dissipative environment with a relaxation rate of $T_1 = \SI{1.7}{\micro\second}$ and coherence time of $T_2 = \SI{0.6}{\micro\second}$. Compared to the single qubit and bright state decay this corresponds to a symmetry protection of a factor 320 and 1300, respectively. We demonstrate coherent control by engineering a drive that allows us to excite the dark state and directly observe the phase dependence. Energies and symmetries of the collective states in the first and second excitation manifold are captured by a phase-sensitive spectroscopy. Our experiments demonstrate a step towards the realization of quantum computation with decoherence-free subspaces~\cite{paulisch_universal_2016-2}. Replacing the copper of the waveguide with a superconductor would reduce unwanted dissipation to values as low as \SI{1}{\decibel \per \kilo \meter}~\cite{magnard_microwave_2020} and enable the efficient generation of collective states to study quantum state transfer over larger distances~\cite{vermersch_quantum_2017,xiang_intracity_2017,vogell_deterministic_2017}. Furthermore, scaling our systems to several 10's of qubits opens up the possibility to study the dynamics of interacting quantum many-body systems in an open environment~\cite{gonzalez-tudela_deterministic_2015,albrecht_subradiant_2019}.


\textbf{Data availability}
The data that support the findings of this study are available on Zenodo.
\newline
\textbf{Code availability}
The code used for the data analysis and simulated results is available from the corresponding author upon reasonable request.
\newline
\textbf{Acknowledgments}
We thank Andreas Strasser for fabricating the waveguide sample. We would like to thank Eric I. Rosenthal for valuable comments on the manuscript. M.Z. and S.O. acknowledge funding by the European Research Council (ERC) under the European Unions Horizon 2020 research and innovation program (714235). M.Z. and C.M.F.S. acknowledge support by the Austrian Science Fund FWF within the DK-ALM (W1259-N27). R.A. acknowledges support from the Austrian Science Fund FWF within the SFB-BeyondC (F7106-N38). T.O. and M.S. acknowledge funding by the Emil Aaltonen Foundation and by the Academy of Finland (316619, 320086).
\newline
\textbf{Competing Interests:} The authors declare no competing interests.
\newline
\textbf{Author  contributions:} M.Z. and G.K. conceived and designed the experiment. M.Z. simulated and fabricated the devices. M.Z. conducted the measurements. M.Z. and C.M.F.S. analyzed the data. T.O. and M.S. developed the theoretical model and performed the simulations. M.Z. and G.K. wrote the manuscript. All authors discussed the results and contributed to the writing of the manuscript.
\newline
\textbf{Correspondence and requests for materials} should be addressed to M.Z. or G.K.
\bibliography{references}

\begin{thebibliography}{43}%
\makeatletter
\providecommand \@ifxundefined [1]{%
 \@ifx{#1\undefined}
}%
\providecommand \@ifnum [1]{%
 \ifnum #1\expandafter \@firstoftwo
 \else \expandafter \@secondoftwo
 \fi
}%
\providecommand \@ifx [1]{%
 \ifx #1\expandafter \@firstoftwo
 \else \expandafter \@secondoftwo
 \fi
}%
\providecommand \natexlab [1]{#1}%
\providecommand \enquote  [1]{``#1''}%
\providecommand \bibnamefont  [1]{#1}%
\providecommand \bibfnamefont [1]{#1}%
\providecommand \citenamefont [1]{#1}%
\providecommand \href@noop [0]{\@secondoftwo}%
\providecommand \href [0]{\begingroup \@sanitize@url \@href}%
\providecommand \@href[1]{\@@startlink{#1}\@@href}%
\providecommand \@@href[1]{\endgroup#1\@@endlink}%
\providecommand \@sanitize@url [0]{\catcode `\\12\catcode `\$12\catcode
  `\&12\catcode `\#12\catcode `\^12\catcode `\_12\catcode `\%12\relax}%
\providecommand \@@startlink[1]{}%
\providecommand \@@endlink[0]{}%
\providecommand \url  [0]{\begingroup\@sanitize@url \@url }%
\providecommand \@url [1]{\endgroup\@href {#1}{\urlprefix }}%
\providecommand \urlprefix  [0]{URL }%
\providecommand \Eprint [0]{\href }%
\providecommand \doibase [0]{https://doi.org/}%
\providecommand \selectlanguage [0]{\@gobble}%
\providecommand \bibinfo  [0]{\@secondoftwo}%
\providecommand \bibfield  [0]{\@secondoftwo}%
\providecommand \translation [1]{[#1]}%
\providecommand \BibitemOpen [0]{}%
\providecommand \bibitemStop [0]{}%
\providecommand \bibitemNoStop [0]{.\EOS\space}%
\providecommand \EOS [0]{\spacefactor3000\relax}%
\providecommand \BibitemShut  [1]{\csname bibitem#1\endcsname}%
\let\auto@bib@innerbib\@empty
\bibitem [{\citenamefont {Roy}\ \emph {et~al.}(2017)\citenamefont {Roy},
  \citenamefont {Wilson},\ and\ \citenamefont
  {Firstenberg}}]{roy_colloquium_2017}%
  \BibitemOpen
  \bibfield  {author} {\bibinfo {author} {\bibfnamefont {D.}~\bibnamefont
  {Roy}}, \bibinfo {author} {\bibfnamefont {C.~M.}\ \bibnamefont {Wilson}},\
  and\ \bibinfo {author} {\bibfnamefont {O.}~\bibnamefont {Firstenberg}},\
  }\bibfield  {title} {\bibinfo {title} {Colloquium: {{Strongly}} interacting
  photons in one-dimensional continuum},\ }\href
  {https://doi.org/10.1103/RevModPhys.89.021001} {\bibfield  {journal}
  {\bibinfo  {journal} {Reviews of Modern Physics}\ }\textbf {\bibinfo {volume}
  {89}},\ \bibinfo {pages} {021001} (\bibinfo {year} {2017})}\BibitemShut
  {NoStop}%
\bibitem [{\citenamefont {Masson}\ and\ \citenamefont
  {{Asenjo-Garcia}}(2020)}]{masson_atomic-waveguide_2020}%
  \BibitemOpen
  \bibfield  {author} {\bibinfo {author} {\bibfnamefont {S.~J.}\ \bibnamefont
  {Masson}}\ and\ \bibinfo {author} {\bibfnamefont {A.}~\bibnamefont
  {{Asenjo-Garcia}}},\ }\bibfield  {title} {\bibinfo {title} {Atomic-waveguide
  quantum electrodynamics},\ }\href {https://doi.org/10/gh2n43} {\bibfield
  {journal} {\bibinfo  {journal} {Physical Review Research}\ }\textbf {\bibinfo
  {volume} {2}},\ \bibinfo {pages} {043213} (\bibinfo {year}
  {2020})}\BibitemShut {NoStop}%
\bibitem [{\citenamefont {Goban}\ \emph {et~al.}(2014)\citenamefont {Goban},
  \citenamefont {Hung}, \citenamefont {Yu}, \citenamefont {Hood}, \citenamefont
  {Muniz}, \citenamefont {Lee}, \citenamefont {Martin}, \citenamefont
  {McClung}, \citenamefont {Choi}, \citenamefont {Chang}, \citenamefont
  {Painter},\ and\ \citenamefont {Kimble}}]{goban_atomlight_2014}%
  \BibitemOpen
  \bibfield  {author} {\bibinfo {author} {\bibfnamefont {A.}~\bibnamefont
  {Goban}}, \bibinfo {author} {\bibfnamefont {C.-L.}\ \bibnamefont {Hung}},
  \bibinfo {author} {\bibfnamefont {S.-P.}\ \bibnamefont {Yu}}, \bibinfo
  {author} {\bibfnamefont {J.}~\bibnamefont {Hood}}, \bibinfo {author}
  {\bibfnamefont {J.}~\bibnamefont {Muniz}}, \bibinfo {author} {\bibfnamefont
  {J.}~\bibnamefont {Lee}}, \bibinfo {author} {\bibfnamefont {M.}~\bibnamefont
  {Martin}}, \bibinfo {author} {\bibfnamefont {A.}~\bibnamefont {McClung}},
  \bibinfo {author} {\bibfnamefont {K.}~\bibnamefont {Choi}}, \bibinfo {author}
  {\bibfnamefont {D.}~\bibnamefont {Chang}}, \bibinfo {author} {\bibfnamefont
  {O.}~\bibnamefont {Painter}},\ and\ \bibinfo {author} {\bibfnamefont
  {H.}~\bibnamefont {Kimble}},\ }\bibfield  {title} {\bibinfo {title}
  {Atom\textendash light interactions in photonic crystals},\ }\href
  {https://doi.org/10.1038/ncomms4808} {\bibfield  {journal} {\bibinfo
  {journal} {Nature Communications}\ }\textbf {\bibinfo {volume} {5}},\
  \bibinfo {pages} {3808} (\bibinfo {year} {2014})}\BibitemShut {NoStop}%
\bibitem [{\citenamefont {Corzo}\ \emph {et~al.}(2016)\citenamefont {Corzo},
  \citenamefont {Gouraud}, \citenamefont {Chandra}, \citenamefont {Goban},
  \citenamefont {Sheremet}, \citenamefont {Kupriyanov},\ and\ \citenamefont
  {Laurat}}]{corzo_large_2016}%
  \BibitemOpen
  \bibfield  {author} {\bibinfo {author} {\bibfnamefont {N.~V.}\ \bibnamefont
  {Corzo}}, \bibinfo {author} {\bibfnamefont {B.}~\bibnamefont {Gouraud}},
  \bibinfo {author} {\bibfnamefont {A.}~\bibnamefont {Chandra}}, \bibinfo
  {author} {\bibfnamefont {A.}~\bibnamefont {Goban}}, \bibinfo {author}
  {\bibfnamefont {A.~S.}\ \bibnamefont {Sheremet}}, \bibinfo {author}
  {\bibfnamefont {D.~V.}\ \bibnamefont {Kupriyanov}},\ and\ \bibinfo {author}
  {\bibfnamefont {J.}~\bibnamefont {Laurat}},\ }\bibfield  {title} {\bibinfo
  {title} {Large {{Bragg Reflection}} from {{One}}-{{Dimensional Chains}} of
  {{Trapped Atoms Near}} a {{Nanoscale Waveguide}}},\ }\href
  {https://doi.org/10.1103/PhysRevLett.117.133603} {\bibfield  {journal}
  {\bibinfo  {journal} {Physical Review Letters}\ }\textbf {\bibinfo {volume}
  {117}},\ \bibinfo {pages} {133603} (\bibinfo {year} {2016})}\BibitemShut
  {NoStop}%
\bibitem [{\citenamefont {Lodahl}\ \emph {et~al.}(2015)\citenamefont {Lodahl},
  \citenamefont {Mahmoodian},\ and\ \citenamefont
  {Stobbe}}]{lodahl_interfacing_2015}%
  \BibitemOpen
  \bibfield  {author} {\bibinfo {author} {\bibfnamefont {P.}~\bibnamefont
  {Lodahl}}, \bibinfo {author} {\bibfnamefont {S.}~\bibnamefont {Mahmoodian}},\
  and\ \bibinfo {author} {\bibfnamefont {S.}~\bibnamefont {Stobbe}},\
  }\bibfield  {title} {\bibinfo {title} {Interfacing single photons and single
  quantum dots with photonic nanostructures},\ }\href
  {https://doi.org/10.1103/RevModPhys.87.347} {\bibfield  {journal} {\bibinfo
  {journal} {Reviews of Modern Physics}\ }\textbf {\bibinfo {volume} {87}},\
  \bibinfo {pages} {347} (\bibinfo {year} {2015})}\BibitemShut {NoStop}%
\bibitem [{\citenamefont {Kannan}\ \emph
  {et~al.}(2020{\natexlab{a}})\citenamefont {Kannan}, \citenamefont
  {Ruckriegel}, \citenamefont {Campbell}, \citenamefont {Frisk~Kockum},
  \citenamefont {Braum{\"u}ller}, \citenamefont {Kim}, \citenamefont
  {Kjaergaard}, \citenamefont {Krantz}, \citenamefont {Melville}, \citenamefont
  {Niedzielski}, \citenamefont {Veps{\"a}l{\"a}inen}, \citenamefont {Winik},
  \citenamefont {Yoder}, \citenamefont {Nori}, \citenamefont {Orlando},
  \citenamefont {Gustavsson},\ and\ \citenamefont
  {Oliver}}]{kannan_waveguide_2020}%
  \BibitemOpen
  \bibfield  {author} {\bibinfo {author} {\bibfnamefont {B.}~\bibnamefont
  {Kannan}}, \bibinfo {author} {\bibfnamefont {M.~J.}\ \bibnamefont
  {Ruckriegel}}, \bibinfo {author} {\bibfnamefont {D.~L.}\ \bibnamefont
  {Campbell}}, \bibinfo {author} {\bibfnamefont {A.}~\bibnamefont
  {Frisk~Kockum}}, \bibinfo {author} {\bibfnamefont {J.}~\bibnamefont
  {Braum{\"u}ller}}, \bibinfo {author} {\bibfnamefont {D.~K.}\ \bibnamefont
  {Kim}}, \bibinfo {author} {\bibfnamefont {M.}~\bibnamefont {Kjaergaard}},
  \bibinfo {author} {\bibfnamefont {P.}~\bibnamefont {Krantz}}, \bibinfo
  {author} {\bibfnamefont {A.}~\bibnamefont {Melville}}, \bibinfo {author}
  {\bibfnamefont {B.~M.}\ \bibnamefont {Niedzielski}}, \bibinfo {author}
  {\bibfnamefont {A.}~\bibnamefont {Veps{\"a}l{\"a}inen}}, \bibinfo {author}
  {\bibfnamefont {R.}~\bibnamefont {Winik}}, \bibinfo {author} {\bibfnamefont
  {J.~L.}\ \bibnamefont {Yoder}}, \bibinfo {author} {\bibfnamefont
  {F.}~\bibnamefont {Nori}}, \bibinfo {author} {\bibfnamefont {T.~P.}\
  \bibnamefont {Orlando}}, \bibinfo {author} {\bibfnamefont {S.}~\bibnamefont
  {Gustavsson}},\ and\ \bibinfo {author} {\bibfnamefont {W.~D.}\ \bibnamefont
  {Oliver}},\ }\bibfield  {title} {\bibinfo {title} {Waveguide quantum
  electrodynamics with superconducting artificial giant atoms},\ }\href
  {https://doi.org/10.1038/s41586-020-2529-9} {\bibfield  {journal} {\bibinfo
  {journal} {Nature}\ }\textbf {\bibinfo {volume} {583}},\ \bibinfo {pages}
  {775} (\bibinfo {year} {2020}{\natexlab{a}})}\BibitemShut {NoStop}%
\bibitem [{\citenamefont {Astafiev}\ \emph {et~al.}(2010)\citenamefont
  {Astafiev}, \citenamefont {Zagoskin}, \citenamefont {Abdumalikov},
  \citenamefont {Pashkin}, \citenamefont {Yamamoto}, \citenamefont {Inomata},
  \citenamefont {Nakamura},\ and\ \citenamefont
  {Tsai}}]{astafiev_resonance_2010-2}%
  \BibitemOpen
  \bibfield  {author} {\bibinfo {author} {\bibfnamefont {O.}~\bibnamefont
  {Astafiev}}, \bibinfo {author} {\bibfnamefont {A.~M.}\ \bibnamefont
  {Zagoskin}}, \bibinfo {author} {\bibfnamefont {A.~A.}\ \bibnamefont
  {Abdumalikov}}, \bibinfo {author} {\bibfnamefont {Y.~A.}\ \bibnamefont
  {Pashkin}}, \bibinfo {author} {\bibfnamefont {T.}~\bibnamefont {Yamamoto}},
  \bibinfo {author} {\bibfnamefont {K.}~\bibnamefont {Inomata}}, \bibinfo
  {author} {\bibfnamefont {Y.}~\bibnamefont {Nakamura}},\ and\ \bibinfo
  {author} {\bibfnamefont {J.~S.}\ \bibnamefont {Tsai}},\ }\bibfield  {title}
  {\bibinfo {title} {Resonance {{Fluorescence}} of a {{Single Artificial
  Atom}}},\ }\href {https://doi.org/10.1126/science.1181918} {\bibfield
  {journal} {\bibinfo  {journal} {Science}\ }\textbf {\bibinfo {volume}
  {327}},\ \bibinfo {pages} {840} (\bibinfo {year} {2010})}\BibitemShut
  {NoStop}%
\bibitem [{\citenamefont {Baur}\ \emph {et~al.}(2009)\citenamefont {Baur},
  \citenamefont {Filipp}, \citenamefont {Bianchetti}, \citenamefont {Fink},
  \citenamefont {G{\"o}ppl}, \citenamefont {Steffen}, \citenamefont {Leek},
  \citenamefont {Blais},\ and\ \citenamefont
  {Wallraff}}]{baur_measurement_2009}%
  \BibitemOpen
  \bibfield  {author} {\bibinfo {author} {\bibfnamefont {M.}~\bibnamefont
  {Baur}}, \bibinfo {author} {\bibfnamefont {S.}~\bibnamefont {Filipp}},
  \bibinfo {author} {\bibfnamefont {R.}~\bibnamefont {Bianchetti}}, \bibinfo
  {author} {\bibfnamefont {J.~M.}\ \bibnamefont {Fink}}, \bibinfo {author}
  {\bibfnamefont {M.}~\bibnamefont {G{\"o}ppl}}, \bibinfo {author}
  {\bibfnamefont {L.}~\bibnamefont {Steffen}}, \bibinfo {author} {\bibfnamefont
  {P.~J.}\ \bibnamefont {Leek}}, \bibinfo {author} {\bibfnamefont
  {A.}~\bibnamefont {Blais}},\ and\ \bibinfo {author} {\bibfnamefont
  {A.}~\bibnamefont {Wallraff}},\ }\bibfield  {title} {\bibinfo {title}
  {Measurement of {{Autler}}-{{Townes}} and {{Mollow Transitions}} in a
  {{Strongly Driven Superconducting Qubit}}},\ }\href
  {https://doi.org/10.1103/PhysRevLett.102.243602} {\bibfield  {journal}
  {\bibinfo  {journal} {Physical Review Letters}\ }\textbf {\bibinfo {volume}
  {102}},\ \bibinfo {pages} {243602} (\bibinfo {year} {2009})}\BibitemShut
  {NoStop}%
\bibitem [{\citenamefont {{Forn-D{\'i}az}}\ \emph {et~al.}(2017)\citenamefont
  {{Forn-D{\'i}az}}, \citenamefont {{Garc{\'i}a-Ripoll}}, \citenamefont
  {Peropadre}, \citenamefont {Orgiazzi}, \citenamefont {Yurtalan},
  \citenamefont {Belyansky}, \citenamefont {Wilson},\ and\ \citenamefont
  {Lupascu}}]{forn-diaz_ultrastrong_2017}%
  \BibitemOpen
  \bibfield  {author} {\bibinfo {author} {\bibfnamefont {P.}~\bibnamefont
  {{Forn-D{\'i}az}}}, \bibinfo {author} {\bibfnamefont {J.~J.}\ \bibnamefont
  {{Garc{\'i}a-Ripoll}}}, \bibinfo {author} {\bibfnamefont {B.}~\bibnamefont
  {Peropadre}}, \bibinfo {author} {\bibfnamefont {J.-L.}\ \bibnamefont
  {Orgiazzi}}, \bibinfo {author} {\bibfnamefont {M.~A.}\ \bibnamefont
  {Yurtalan}}, \bibinfo {author} {\bibfnamefont {R.}~\bibnamefont {Belyansky}},
  \bibinfo {author} {\bibfnamefont {C.~M.}\ \bibnamefont {Wilson}},\ and\
  \bibinfo {author} {\bibfnamefont {A.}~\bibnamefont {Lupascu}},\ }\bibfield
  {title} {\bibinfo {title} {Ultrastrong coupling of a single artificial atom
  to an electromagnetic continuum in the nonperturbative regime},\ }\href
  {https://doi.org/10.1038/nphys3905} {\bibfield  {journal} {\bibinfo
  {journal} {Nature Physics}\ }\textbf {\bibinfo {volume} {13}},\ \bibinfo
  {pages} {39} (\bibinfo {year} {2017})}\BibitemShut {NoStop}%
\bibitem [{\citenamefont {Hoi}\ \emph {et~al.}(2012)\citenamefont {Hoi},
  \citenamefont {Palomaki}, \citenamefont {Lindkvist}, \citenamefont
  {Johansson}, \citenamefont {Delsing},\ and\ \citenamefont
  {Wilson}}]{hoi_generation_2012}%
  \BibitemOpen
  \bibfield  {author} {\bibinfo {author} {\bibfnamefont {I.-C.}\ \bibnamefont
  {Hoi}}, \bibinfo {author} {\bibfnamefont {T.}~\bibnamefont {Palomaki}},
  \bibinfo {author} {\bibfnamefont {J.}~\bibnamefont {Lindkvist}}, \bibinfo
  {author} {\bibfnamefont {G.}~\bibnamefont {Johansson}}, \bibinfo {author}
  {\bibfnamefont {P.}~\bibnamefont {Delsing}},\ and\ \bibinfo {author}
  {\bibfnamefont {C.~M.}\ \bibnamefont {Wilson}},\ }\bibfield  {title}
  {\bibinfo {title} {Generation of {{Nonclassical Microwave States Using}} an
  {{Artificial Atom}} in {{1D Open Space}}},\ }\href
  {https://doi.org/10.1103/PhysRevLett.108.263601} {\bibfield  {journal}
  {\bibinfo  {journal} {Physical Review Letters}\ }\textbf {\bibinfo {volume}
  {108}},\ \bibinfo {pages} {263601} (\bibinfo {year} {2012})}\BibitemShut
  {NoStop}%
\bibitem [{\citenamefont {Kannan}\ \emph
  {et~al.}(2020{\natexlab{b}})\citenamefont {Kannan}, \citenamefont {Campbell},
  \citenamefont {Vasconcelos}, \citenamefont {Winik}, \citenamefont {Kim},
  \citenamefont {Kjaergaard}, \citenamefont {Krantz}, \citenamefont {Melville},
  \citenamefont {Niedzielski}, \citenamefont {Yoder}, \citenamefont {Orlando},
  \citenamefont {Gustavsson},\ and\ \citenamefont
  {Oliver}}]{kannan_generating_2020-1}%
  \BibitemOpen
  \bibfield  {author} {\bibinfo {author} {\bibfnamefont {B.}~\bibnamefont
  {Kannan}}, \bibinfo {author} {\bibfnamefont {D.~L.}\ \bibnamefont
  {Campbell}}, \bibinfo {author} {\bibfnamefont {F.}~\bibnamefont
  {Vasconcelos}}, \bibinfo {author} {\bibfnamefont {R.}~\bibnamefont {Winik}},
  \bibinfo {author} {\bibfnamefont {D.~K.}\ \bibnamefont {Kim}}, \bibinfo
  {author} {\bibfnamefont {M.}~\bibnamefont {Kjaergaard}}, \bibinfo {author}
  {\bibfnamefont {P.}~\bibnamefont {Krantz}}, \bibinfo {author} {\bibfnamefont
  {A.}~\bibnamefont {Melville}}, \bibinfo {author} {\bibfnamefont {B.~M.}\
  \bibnamefont {Niedzielski}}, \bibinfo {author} {\bibfnamefont {J.~L.}\
  \bibnamefont {Yoder}}, \bibinfo {author} {\bibfnamefont {T.~P.}\ \bibnamefont
  {Orlando}}, \bibinfo {author} {\bibfnamefont {S.}~\bibnamefont
  {Gustavsson}},\ and\ \bibinfo {author} {\bibfnamefont {W.~D.}\ \bibnamefont
  {Oliver}},\ }\bibfield  {title} {\bibinfo {title} {Generating spatially
  entangled itinerant photons with waveguide quantum electrodynamics},\ }\href
  {https://doi.org/10.1126/sciadv.abb8780} {\bibfield  {journal} {\bibinfo
  {journal} {Science Advances}\ }\textbf {\bibinfo {volume} {6}},\ \bibinfo
  {pages} {eabb8780} (\bibinfo {year} {2020}{\natexlab{b}})}\BibitemShut
  {NoStop}%
\bibitem [{\citenamefont {Sundaresan}\ \emph {et~al.}(2019)\citenamefont
  {Sundaresan}, \citenamefont {Lundgren}, \citenamefont {Zhu}, \citenamefont
  {Gorshkov},\ and\ \citenamefont {Houck}}]{sundaresan_interacting_2019}%
  \BibitemOpen
  \bibfield  {author} {\bibinfo {author} {\bibfnamefont {N.~M.}\ \bibnamefont
  {Sundaresan}}, \bibinfo {author} {\bibfnamefont {R.}~\bibnamefont
  {Lundgren}}, \bibinfo {author} {\bibfnamefont {G.}~\bibnamefont {Zhu}},
  \bibinfo {author} {\bibfnamefont {A.~V.}\ \bibnamefont {Gorshkov}},\ and\
  \bibinfo {author} {\bibfnamefont {A.~A.}\ \bibnamefont {Houck}},\ }\bibfield
  {title} {\bibinfo {title} {Interacting {{Qubit}}-{{Photon Bound States}} with
  {{Superconducting Circuits}}},\ }\href
  {https://doi.org/10.1103/PhysRevX.9.011021} {\bibfield  {journal} {\bibinfo
  {journal} {Physical Review X}\ }\textbf {\bibinfo {volume} {9}},\ \bibinfo
  {pages} {011021} (\bibinfo {year} {2019})}\BibitemShut {NoStop}%
\bibitem [{\citenamefont {Lodahl}\ \emph {et~al.}(2017)\citenamefont {Lodahl},
  \citenamefont {Mahmoodian}, \citenamefont {Stobbe}, \citenamefont
  {Rauschenbeutel}, \citenamefont {Schneeweiss}, \citenamefont {Volz},
  \citenamefont {Pichler},\ and\ \citenamefont
  {Zoller}}]{lodahl_chiral_2017-1}%
  \BibitemOpen
  \bibfield  {author} {\bibinfo {author} {\bibfnamefont {P.}~\bibnamefont
  {Lodahl}}, \bibinfo {author} {\bibfnamefont {S.}~\bibnamefont {Mahmoodian}},
  \bibinfo {author} {\bibfnamefont {S.}~\bibnamefont {Stobbe}}, \bibinfo
  {author} {\bibfnamefont {A.}~\bibnamefont {Rauschenbeutel}}, \bibinfo
  {author} {\bibfnamefont {P.}~\bibnamefont {Schneeweiss}}, \bibinfo {author}
  {\bibfnamefont {J.}~\bibnamefont {Volz}}, \bibinfo {author} {\bibfnamefont
  {H.}~\bibnamefont {Pichler}},\ and\ \bibinfo {author} {\bibfnamefont
  {P.}~\bibnamefont {Zoller}},\ }\bibfield  {title} {\bibinfo {title} {Chiral
  quantum optics},\ }\href {https://doi.org/10.1038/nature21037} {\bibfield
  {journal} {\bibinfo  {journal} {Nature}\ }\textbf {\bibinfo {volume} {541}},\
  \bibinfo {pages} {473} (\bibinfo {year} {2017})}\BibitemShut {NoStop}%
\bibitem [{\citenamefont {Kim}\ \emph {et~al.}(2021)\citenamefont {Kim},
  \citenamefont {Zhang}, \citenamefont {Ferreira}, \citenamefont {Banker},
  \citenamefont {Iverson}, \citenamefont {Sipahigil}, \citenamefont {Bello},
  \citenamefont {{Gonz{\'a}lez-Tudela}}, \citenamefont {Mirhosseini},\ and\
  \citenamefont {Painter}}]{kim_quantum_2021}%
  \BibitemOpen
  \bibfield  {author} {\bibinfo {author} {\bibfnamefont {E.}~\bibnamefont
  {Kim}}, \bibinfo {author} {\bibfnamefont {X.}~\bibnamefont {Zhang}}, \bibinfo
  {author} {\bibfnamefont {V.~S.}\ \bibnamefont {Ferreira}}, \bibinfo {author}
  {\bibfnamefont {J.}~\bibnamefont {Banker}}, \bibinfo {author} {\bibfnamefont
  {J.~K.}\ \bibnamefont {Iverson}}, \bibinfo {author} {\bibfnamefont
  {A.}~\bibnamefont {Sipahigil}}, \bibinfo {author} {\bibfnamefont
  {M.}~\bibnamefont {Bello}}, \bibinfo {author} {\bibfnamefont
  {A.}~\bibnamefont {{Gonz{\'a}lez-Tudela}}}, \bibinfo {author} {\bibfnamefont
  {M.}~\bibnamefont {Mirhosseini}},\ and\ \bibinfo {author} {\bibfnamefont
  {O.}~\bibnamefont {Painter}},\ }\bibfield  {title} {\bibinfo {title} {Quantum
  {{Electrodynamics}} in a {{Topological Waveguide}}},\ }\href
  {https://doi.org/10.1103/PhysRevX.11.011015} {\bibfield  {journal} {\bibinfo
  {journal} {Physical Review X}\ }\textbf {\bibinfo {volume} {11}},\ \bibinfo
  {pages} {011015} (\bibinfo {year} {2021})}\BibitemShut {NoStop}%
\bibitem [{\citenamefont {van Loo}\ \emph {et~al.}(2013)\citenamefont {van
  Loo}, \citenamefont {Fedorov}, \citenamefont {Lalumi{\`e}re}, \citenamefont
  {Sanders}, \citenamefont {Blais},\ and\ \citenamefont
  {Wallraff}}]{loo_photon-mediated_2013}%
  \BibitemOpen
  \bibfield  {author} {\bibinfo {author} {\bibfnamefont {A.~F.}\ \bibnamefont
  {van Loo}}, \bibinfo {author} {\bibfnamefont {A.}~\bibnamefont {Fedorov}},
  \bibinfo {author} {\bibfnamefont {K.}~\bibnamefont {Lalumi{\`e}re}}, \bibinfo
  {author} {\bibfnamefont {B.~C.}\ \bibnamefont {Sanders}}, \bibinfo {author}
  {\bibfnamefont {A.}~\bibnamefont {Blais}},\ and\ \bibinfo {author}
  {\bibfnamefont {A.}~\bibnamefont {Wallraff}},\ }\bibfield  {title} {\bibinfo
  {title} {Photon-{{Mediated Interactions Between Distant Artificial Atoms}}},\
  }\href {https://doi.org/10.1126/science.1244324} {\bibfield  {journal}
  {\bibinfo  {journal} {Science}\ ,\ \bibinfo {pages} {1244324}} (\bibinfo
  {year} {2013})}\BibitemShut {NoStop}%
\bibitem [{\citenamefont {Lalumi{\`e}re}\ \emph {et~al.}(2013)\citenamefont
  {Lalumi{\`e}re}, \citenamefont {Sanders}, \citenamefont {{van Loo}},
  \citenamefont {Fedorov}, \citenamefont {Wallraff},\ and\ \citenamefont
  {Blais}}]{lalumiere_input-output_2013-2}%
  \BibitemOpen
  \bibfield  {author} {\bibinfo {author} {\bibfnamefont {K.}~\bibnamefont
  {Lalumi{\`e}re}}, \bibinfo {author} {\bibfnamefont {B.~C.}\ \bibnamefont
  {Sanders}}, \bibinfo {author} {\bibfnamefont {A.~F.}\ \bibnamefont {{van
  Loo}}}, \bibinfo {author} {\bibfnamefont {A.}~\bibnamefont {Fedorov}},
  \bibinfo {author} {\bibfnamefont {A.}~\bibnamefont {Wallraff}},\ and\
  \bibinfo {author} {\bibfnamefont {A.}~\bibnamefont {Blais}},\ }\bibfield
  {title} {\bibinfo {title} {Input-output theory for waveguide {{QED}} with an
  ensemble of inhomogeneous atoms},\ }\href
  {https://doi.org/10.1103/PhysRevA.88.043806} {\bibfield  {journal} {\bibinfo
  {journal} {Physical Review A}\ }\textbf {\bibinfo {volume} {88}},\ \bibinfo
  {pages} {043806} (\bibinfo {year} {2013})}\BibitemShut {NoStop}%
\bibitem [{\citenamefont {Kockum}\ \emph {et~al.}(2018)\citenamefont {Kockum},
  \citenamefont {Johansson},\ and\ \citenamefont
  {Nori}}]{kockum_decoherence-free_2018}%
  \BibitemOpen
  \bibfield  {author} {\bibinfo {author} {\bibfnamefont {A.~F.}\ \bibnamefont
  {Kockum}}, \bibinfo {author} {\bibfnamefont {G.}~\bibnamefont {Johansson}},\
  and\ \bibinfo {author} {\bibfnamefont {F.}~\bibnamefont {Nori}},\ }\bibfield
  {title} {\bibinfo {title} {Decoherence-{{Free Interaction}} between {{Giant
  Atoms}} in {{Waveguide Quantum Electrodynamics}}},\ }\href
  {https://doi.org/10.1103/PhysRevLett.120.140404} {\bibfield  {journal}
  {\bibinfo  {journal} {Physical Review Letters}\ }\textbf {\bibinfo {volume}
  {120}},\ \bibinfo {pages} {140404} (\bibinfo {year} {2018})}\BibitemShut
  {NoStop}%
\bibitem [{\citenamefont {Monz}\ \emph {et~al.}(2009)\citenamefont {Monz},
  \citenamefont {Kim}, \citenamefont {Villar}, \citenamefont {Schindler},
  \citenamefont {Chwalla}, \citenamefont {Riebe}, \citenamefont {Roos},
  \citenamefont {H{\"a}ffner}, \citenamefont {H{\"a}nsel}, \citenamefont
  {Hennrich},\ and\ \citenamefont {Blatt}}]{monz_realization_2009}%
  \BibitemOpen
  \bibfield  {author} {\bibinfo {author} {\bibfnamefont {T.}~\bibnamefont
  {Monz}}, \bibinfo {author} {\bibfnamefont {K.}~\bibnamefont {Kim}}, \bibinfo
  {author} {\bibfnamefont {A.~S.}\ \bibnamefont {Villar}}, \bibinfo {author}
  {\bibfnamefont {P.}~\bibnamefont {Schindler}}, \bibinfo {author}
  {\bibfnamefont {M.}~\bibnamefont {Chwalla}}, \bibinfo {author} {\bibfnamefont
  {M.}~\bibnamefont {Riebe}}, \bibinfo {author} {\bibfnamefont {C.~F.}\
  \bibnamefont {Roos}}, \bibinfo {author} {\bibfnamefont {H.}~\bibnamefont
  {H{\"a}ffner}}, \bibinfo {author} {\bibfnamefont {W.}~\bibnamefont
  {H{\"a}nsel}}, \bibinfo {author} {\bibfnamefont {M.}~\bibnamefont
  {Hennrich}},\ and\ \bibinfo {author} {\bibfnamefont {R.}~\bibnamefont
  {Blatt}},\ }\bibfield  {title} {\bibinfo {title} {Realization of {{Universal
  Ion}}-{{Trap Quantum Computation}} with {{Decoherence}}-{{Free Qubits}}},\
  }\href {https://doi.org/10.1103/PhysRevLett.103.200503} {\bibfield  {journal}
  {\bibinfo  {journal} {Physical Review Letters}\ }\textbf {\bibinfo {volume}
  {103}},\ \bibinfo {pages} {200503} (\bibinfo {year} {2009})}\BibitemShut
  {NoStop}%
\bibitem [{\citenamefont {Kwiat}\ \emph {et~al.}(2000)\citenamefont {Kwiat},
  \citenamefont {Berglund}, \citenamefont {Altepeter},\ and\ \citenamefont
  {White}}]{kwiat_experimental_2000}%
  \BibitemOpen
  \bibfield  {author} {\bibinfo {author} {\bibfnamefont {P.~G.}\ \bibnamefont
  {Kwiat}}, \bibinfo {author} {\bibfnamefont {A.~J.}\ \bibnamefont {Berglund}},
  \bibinfo {author} {\bibfnamefont {J.~B.}\ \bibnamefont {Altepeter}},\ and\
  \bibinfo {author} {\bibfnamefont {A.~G.}\ \bibnamefont {White}},\ }\bibfield
  {title} {\bibinfo {title} {Experimental {{Verification}} of
  {{Decoherence}}-{{Free Subspaces}}},\ }\href
  {https://doi.org/10.1126/science.290.5491.498} {\bibfield  {journal}
  {\bibinfo  {journal} {Science}\ }\textbf {\bibinfo {volume} {290}},\ \bibinfo
  {pages} {498} (\bibinfo {year} {2000})}\BibitemShut {NoStop}%
\bibitem [{\citenamefont {Kielpinski}\ \emph {et~al.}(2001)\citenamefont
  {Kielpinski}, \citenamefont {Meyer}, \citenamefont {Rowe}, \citenamefont
  {Sackett}, \citenamefont {Itano}, \citenamefont {Monroe},\ and\ \citenamefont
  {Wineland}}]{kielpinski_decoherence-free_2001}%
  \BibitemOpen
  \bibfield  {author} {\bibinfo {author} {\bibfnamefont {D.}~\bibnamefont
  {Kielpinski}}, \bibinfo {author} {\bibfnamefont {V.}~\bibnamefont {Meyer}},
  \bibinfo {author} {\bibfnamefont {M.~A.}\ \bibnamefont {Rowe}}, \bibinfo
  {author} {\bibfnamefont {C.~A.}\ \bibnamefont {Sackett}}, \bibinfo {author}
  {\bibfnamefont {W.~M.}\ \bibnamefont {Itano}}, \bibinfo {author}
  {\bibfnamefont {C.}~\bibnamefont {Monroe}},\ and\ \bibinfo {author}
  {\bibfnamefont {D.~J.}\ \bibnamefont {Wineland}},\ }\bibfield  {title}
  {\bibinfo {title} {A {{Decoherence}}-{{Free Quantum Memory Using Trapped
  Ions}}},\ }\href {https://doi.org/10.1126/science.1057357} {\bibfield
  {journal} {\bibinfo  {journal} {Science}\ }\textbf {\bibinfo {volume}
  {291}},\ \bibinfo {pages} {1013} (\bibinfo {year} {2001})}\BibitemShut
  {NoStop}%
\bibitem [{\citenamefont {Lidar}\ \emph {et~al.}(1998)\citenamefont {Lidar},
  \citenamefont {Chuang},\ and\ \citenamefont
  {Whaley}}]{lidar_decoherence-free_1998}%
  \BibitemOpen
  \bibfield  {author} {\bibinfo {author} {\bibfnamefont {D.~A.}\ \bibnamefont
  {Lidar}}, \bibinfo {author} {\bibfnamefont {I.~L.}\ \bibnamefont {Chuang}},\
  and\ \bibinfo {author} {\bibfnamefont {K.~B.}\ \bibnamefont {Whaley}},\
  }\bibfield  {title} {\bibinfo {title} {Decoherence-{{Free Subspaces}} for
  {{Quantum Computation}}},\ }\href
  {https://doi.org/10.1103/PhysRevLett.81.2594} {\bibfield  {journal} {\bibinfo
   {journal} {Physical Review Letters}\ }\textbf {\bibinfo {volume} {81}},\
  \bibinfo {pages} {2594} (\bibinfo {year} {1998})}\BibitemShut {NoStop}%
\bibitem [{\citenamefont {Paulisch}\ \emph {et~al.}(2016)\citenamefont
  {Paulisch}, \citenamefont {Kimble},\ and\ \citenamefont
  {{Gonz{\'a}lez-Tudela}}}]{paulisch_universal_2016-2}%
  \BibitemOpen
  \bibfield  {author} {\bibinfo {author} {\bibfnamefont {V.}~\bibnamefont
  {Paulisch}}, \bibinfo {author} {\bibfnamefont {H.~J.}\ \bibnamefont
  {Kimble}},\ and\ \bibinfo {author} {\bibfnamefont {A.}~\bibnamefont
  {{Gonz{\'a}lez-Tudela}}},\ }\bibfield  {title} {\bibinfo {title} {Universal
  quantum computation in waveguide {{QED}} using decoherence free subspaces},\
  }\href {https://doi.org/10/gf68vr} {\bibfield  {journal} {\bibinfo  {journal}
  {New Journal of Physics}\ }\textbf {\bibinfo {volume} {18}},\ \bibinfo
  {pages} {043041} (\bibinfo {year} {2016})}\BibitemShut {NoStop}%
\bibitem [{\citenamefont {Mlynek}\ \emph {et~al.}(2014)\citenamefont {Mlynek},
  \citenamefont {Abdumalikov}, \citenamefont {Eichler},\ and\ \citenamefont
  {Wallraff}}]{mlynek_observation_2014-3}%
  \BibitemOpen
  \bibfield  {author} {\bibinfo {author} {\bibfnamefont {J.~A.}\ \bibnamefont
  {Mlynek}}, \bibinfo {author} {\bibfnamefont {A.~A.}\ \bibnamefont
  {Abdumalikov}}, \bibinfo {author} {\bibfnamefont {C.}~\bibnamefont
  {Eichler}},\ and\ \bibinfo {author} {\bibfnamefont {A.}~\bibnamefont
  {Wallraff}},\ }\bibfield  {title} {\bibinfo {title} {Observation of {{Dicke}}
  superradiance for two artificial atoms in a cavity with high decay rate},\
  }\href {https://doi.org/10.1038/ncomms6186} {\bibfield  {journal} {\bibinfo
  {journal} {Nature Communications}\ }\textbf {\bibinfo {volume} {5}},\
  \bibinfo {pages} {5186} (\bibinfo {year} {2014})}\BibitemShut {NoStop}%
\bibitem [{\citenamefont {Rosario~Hamann}\ \emph {et~al.}(2018)\citenamefont
  {Rosario~Hamann}, \citenamefont {M{\"u}ller}, \citenamefont {Jerger},
  \citenamefont {Zanner}, \citenamefont {Combes}, \citenamefont {Pletyukhov},
  \citenamefont {Weides}, \citenamefont {Stace},\ and\ \citenamefont
  {Fedorov}}]{rosario_hamann_nonreciprocity_2018}%
  \BibitemOpen
  \bibfield  {author} {\bibinfo {author} {\bibfnamefont {A.}~\bibnamefont
  {Rosario~Hamann}}, \bibinfo {author} {\bibfnamefont {C.}~\bibnamefont
  {M{\"u}ller}}, \bibinfo {author} {\bibfnamefont {M.}~\bibnamefont {Jerger}},
  \bibinfo {author} {\bibfnamefont {M.}~\bibnamefont {Zanner}}, \bibinfo
  {author} {\bibfnamefont {J.}~\bibnamefont {Combes}}, \bibinfo {author}
  {\bibfnamefont {M.}~\bibnamefont {Pletyukhov}}, \bibinfo {author}
  {\bibfnamefont {M.}~\bibnamefont {Weides}}, \bibinfo {author} {\bibfnamefont
  {T.~M.}\ \bibnamefont {Stace}},\ and\ \bibinfo {author} {\bibfnamefont
  {A.}~\bibnamefont {Fedorov}},\ }\bibfield  {title} {\bibinfo {title}
  {Nonreciprocity {{Realized}} with {{Quantum Nonlinearity}}},\ }\href
  {https://doi.org/10.1103/PhysRevLett.121.123601} {\bibfield  {journal}
  {\bibinfo  {journal} {Physical Review Letters}\ }\textbf {\bibinfo {volume}
  {121}},\ \bibinfo {pages} {123601} (\bibinfo {year} {2018})}\BibitemShut
  {NoStop}%
\bibitem [{\citenamefont {Mirhosseini}\ \emph {et~al.}(2019)\citenamefont
  {Mirhosseini}, \citenamefont {Kim}, \citenamefont {Zhang}, \citenamefont
  {Sipahigil}, \citenamefont {Dieterle}, \citenamefont {Keller}, \citenamefont
  {{Asenjo-Garcia}}, \citenamefont {Chang},\ and\ \citenamefont
  {Painter}}]{mirhosseini_cavity_2019-1}%
  \BibitemOpen
  \bibfield  {author} {\bibinfo {author} {\bibfnamefont {M.}~\bibnamefont
  {Mirhosseini}}, \bibinfo {author} {\bibfnamefont {E.}~\bibnamefont {Kim}},
  \bibinfo {author} {\bibfnamefont {X.}~\bibnamefont {Zhang}}, \bibinfo
  {author} {\bibfnamefont {A.}~\bibnamefont {Sipahigil}}, \bibinfo {author}
  {\bibfnamefont {P.~B.}\ \bibnamefont {Dieterle}}, \bibinfo {author}
  {\bibfnamefont {A.~J.}\ \bibnamefont {Keller}}, \bibinfo {author}
  {\bibfnamefont {A.}~\bibnamefont {{Asenjo-Garcia}}}, \bibinfo {author}
  {\bibfnamefont {D.~E.}\ \bibnamefont {Chang}},\ and\ \bibinfo {author}
  {\bibfnamefont {O.}~\bibnamefont {Painter}},\ }\bibfield  {title} {\bibinfo
  {title} {Cavity quantum electrodynamics with atom-like mirrors},\ }\href
  {https://doi.org/10.1038/s41586-019-1196-1} {\bibfield  {journal} {\bibinfo
  {journal} {Nature}\ }\textbf {\bibinfo {volume} {569}},\ \bibinfo {pages}
  {692} (\bibinfo {year} {2019})}\BibitemShut {NoStop}%
\bibitem [{\citenamefont {Koch}\ \emph {et~al.}(2007)\citenamefont {Koch},
  \citenamefont {Yu}, \citenamefont {Gambetta}, \citenamefont {Houck},
  \citenamefont {Schuster}, \citenamefont {Majer}, \citenamefont {Blais},
  \citenamefont {Devoret}, \citenamefont {Girvin},\ and\ \citenamefont
  {Schoelkopf}}]{koch_charge-insensitive_2007-2}%
  \BibitemOpen
  \bibfield  {author} {\bibinfo {author} {\bibfnamefont {J.}~\bibnamefont
  {Koch}}, \bibinfo {author} {\bibfnamefont {T.~M.}\ \bibnamefont {Yu}},
  \bibinfo {author} {\bibfnamefont {J.}~\bibnamefont {Gambetta}}, \bibinfo
  {author} {\bibfnamefont {A.~A.}\ \bibnamefont {Houck}}, \bibinfo {author}
  {\bibfnamefont {D.~I.}\ \bibnamefont {Schuster}}, \bibinfo {author}
  {\bibfnamefont {J.}~\bibnamefont {Majer}}, \bibinfo {author} {\bibfnamefont
  {A.}~\bibnamefont {Blais}}, \bibinfo {author} {\bibfnamefont {M.~H.}\
  \bibnamefont {Devoret}}, \bibinfo {author} {\bibfnamefont {S.~M.}\
  \bibnamefont {Girvin}},\ and\ \bibinfo {author} {\bibfnamefont {R.~J.}\
  \bibnamefont {Schoelkopf}},\ }\bibfield  {title} {\bibinfo {title}
  {Charge-insensitive qubit design derived from the {{Cooper}} pair box},\
  }\href {https://doi.org/10.1103/PhysRevA.76.042319} {\bibfield  {journal}
  {\bibinfo  {journal} {Physical Review A}\ }\textbf {\bibinfo {volume} {76}},\
  \bibinfo {pages} {042319} (\bibinfo {year} {2007})}\BibitemShut {NoStop}%
\bibitem [{\citenamefont {Pozar}(2012)}]{pozar_microwave_2012}%
  \BibitemOpen
  \bibfield  {author} {\bibinfo {author} {\bibfnamefont {D.~M.}\ \bibnamefont
  {Pozar}},\ }\href@noop {} {\emph {\bibinfo {title} {Microwave
  Engineering}}},\ \bibinfo {edition} {4th}\ ed.\ (\bibinfo  {publisher}
  {{Wiley}},\ \bibinfo {address} {{Hoboken, NJ}},\ \bibinfo {year}
  {2012})\BibitemShut {NoStop}%
\bibitem [{\citenamefont {Dalmonte}\ \emph {et~al.}(2015)\citenamefont
  {Dalmonte}, \citenamefont {Mirzaei}, \citenamefont {Muppalla}, \citenamefont
  {Marcos}, \citenamefont {Zoller},\ and\ \citenamefont
  {Kirchmair}}]{dalmonte_realizing_2015-5}%
  \BibitemOpen
  \bibfield  {author} {\bibinfo {author} {\bibfnamefont {M.}~\bibnamefont
  {Dalmonte}}, \bibinfo {author} {\bibfnamefont {S.~I.}\ \bibnamefont
  {Mirzaei}}, \bibinfo {author} {\bibfnamefont {P.~R.}\ \bibnamefont
  {Muppalla}}, \bibinfo {author} {\bibfnamefont {D.}~\bibnamefont {Marcos}},
  \bibinfo {author} {\bibfnamefont {P.}~\bibnamefont {Zoller}},\ and\ \bibinfo
  {author} {\bibfnamefont {G.}~\bibnamefont {Kirchmair}},\ }\bibfield  {title}
  {\bibinfo {title} {Realizing dipolar spin models with arrays of
  superconducting qubits},\ }\href {https://doi.org/10.1103/PhysRevB.92.174507}
  {\bibfield  {journal} {\bibinfo  {journal} {Physical Review B}\ }\textbf
  {\bibinfo {volume} {92}},\ \bibinfo {pages} {174507} (\bibinfo {year}
  {2015})}\BibitemShut {NoStop}%
\bibitem [{\citenamefont {Probst}\ \emph {et~al.}(2015)\citenamefont {Probst},
  \citenamefont {Song}, \citenamefont {Bushev}, \citenamefont {Ustinov},\ and\
  \citenamefont {Weides}}]{probst_efficient_2015-3}%
  \BibitemOpen
  \bibfield  {author} {\bibinfo {author} {\bibfnamefont {S.}~\bibnamefont
  {Probst}}, \bibinfo {author} {\bibfnamefont {F.~B.}\ \bibnamefont {Song}},
  \bibinfo {author} {\bibfnamefont {P.~A.}\ \bibnamefont {Bushev}}, \bibinfo
  {author} {\bibfnamefont {A.~V.}\ \bibnamefont {Ustinov}},\ and\ \bibinfo
  {author} {\bibfnamefont {M.}~\bibnamefont {Weides}},\ }\bibfield  {title}
  {\bibinfo {title} {Efficient and robust analysis of complex scattering data
  under noise in microwave resonators},\ }\href
  {https://doi.org/10.1063/1.4907935} {\bibfield  {journal} {\bibinfo
  {journal} {Review of Scientific Instruments}\ }\textbf {\bibinfo {volume}
  {86}},\ \bibinfo {pages} {024706} (\bibinfo {year} {2015})}\BibitemShut
  {NoStop}%
\bibitem [{\citenamefont {Leibfried}\ \emph {et~al.}(2003)\citenamefont
  {Leibfried}, \citenamefont {Blatt}, \citenamefont {Monroe},\ and\
  \citenamefont {Wineland}}]{leibfried_quantum_2003}%
  \BibitemOpen
  \bibfield  {author} {\bibinfo {author} {\bibfnamefont {D.}~\bibnamefont
  {Leibfried}}, \bibinfo {author} {\bibfnamefont {R.}~\bibnamefont {Blatt}},
  \bibinfo {author} {\bibfnamefont {C.}~\bibnamefont {Monroe}},\ and\ \bibinfo
  {author} {\bibfnamefont {D.}~\bibnamefont {Wineland}},\ }\bibfield  {title}
  {\bibinfo {title} {Quantum dynamics of single trapped ions},\ }\href
  {https://doi.org/10.1103/RevModPhys.75.281} {\bibfield  {journal} {\bibinfo
  {journal} {Reviews of Modern Physics}\ }\textbf {\bibinfo {volume} {75}},\
  \bibinfo {pages} {281} (\bibinfo {year} {2003})}\BibitemShut {NoStop}%
\bibitem [{\citenamefont {Roushan}\ \emph {et~al.}(2017)\citenamefont
  {Roushan}, \citenamefont {Neill}, \citenamefont {Tangpanitanon},
  \citenamefont {Bastidas}, \citenamefont {Megrant}, \citenamefont {Barends},
  \citenamefont {Chen}, \citenamefont {Chen}, \citenamefont {Chiaro},
  \citenamefont {Dunsworth}, \citenamefont {Fowler}, \citenamefont {Foxen},
  \citenamefont {Giustina}, \citenamefont {Jeffrey}, \citenamefont {Kelly},
  \citenamefont {Lucero}, \citenamefont {Mutus}, \citenamefont {Neeley},
  \citenamefont {Quintana}, \citenamefont {Sank}, \citenamefont {Vainsencher},
  \citenamefont {Wenner}, \citenamefont {White}, \citenamefont {Neven},
  \citenamefont {Angelakis},\ and\ \citenamefont {Martinis}}]{Roushan17}%
  \BibitemOpen
  \bibfield  {author} {\bibinfo {author} {\bibfnamefont {P.}~\bibnamefont
  {Roushan}}, \bibinfo {author} {\bibfnamefont {C.}~\bibnamefont {Neill}},
  \bibinfo {author} {\bibfnamefont {J.}~\bibnamefont {Tangpanitanon}}, \bibinfo
  {author} {\bibfnamefont {V.~M.}\ \bibnamefont {Bastidas}}, \bibinfo {author}
  {\bibfnamefont {A.}~\bibnamefont {Megrant}}, \bibinfo {author} {\bibfnamefont
  {R.}~\bibnamefont {Barends}}, \bibinfo {author} {\bibfnamefont
  {Y.}~\bibnamefont {Chen}}, \bibinfo {author} {\bibfnamefont {Z.}~\bibnamefont
  {Chen}}, \bibinfo {author} {\bibfnamefont {B.}~\bibnamefont {Chiaro}},
  \bibinfo {author} {\bibfnamefont {A.}~\bibnamefont {Dunsworth}}, \bibinfo
  {author} {\bibfnamefont {A.}~\bibnamefont {Fowler}}, \bibinfo {author}
  {\bibfnamefont {B.}~\bibnamefont {Foxen}}, \bibinfo {author} {\bibfnamefont
  {M.}~\bibnamefont {Giustina}}, \bibinfo {author} {\bibfnamefont
  {E.}~\bibnamefont {Jeffrey}}, \bibinfo {author} {\bibfnamefont
  {J.}~\bibnamefont {Kelly}}, \bibinfo {author} {\bibfnamefont
  {E.}~\bibnamefont {Lucero}}, \bibinfo {author} {\bibfnamefont
  {J.}~\bibnamefont {Mutus}}, \bibinfo {author} {\bibfnamefont
  {M.}~\bibnamefont {Neeley}}, \bibinfo {author} {\bibfnamefont
  {C.}~\bibnamefont {Quintana}}, \bibinfo {author} {\bibfnamefont
  {D.}~\bibnamefont {Sank}}, \bibinfo {author} {\bibfnamefont {A.}~\bibnamefont
  {Vainsencher}}, \bibinfo {author} {\bibfnamefont {J.}~\bibnamefont {Wenner}},
  \bibinfo {author} {\bibfnamefont {T.}~\bibnamefont {White}}, \bibinfo
  {author} {\bibfnamefont {H.}~\bibnamefont {Neven}}, \bibinfo {author}
  {\bibfnamefont {D.~G.}\ \bibnamefont {Angelakis}},\ and\ \bibinfo {author}
  {\bibfnamefont {J.}~\bibnamefont {Martinis}},\ }\bibfield  {title} {\bibinfo
  {title} {Spectroscopic signatures of localization with interacting photons in
  superconducting qubits},\ }\href {https://doi.org/10.1126/science.aao1401}
  {\bibfield  {journal} {\bibinfo  {journal} {Science}\ }\textbf {\bibinfo
  {volume} {358}},\ \bibinfo {pages} {1175} (\bibinfo {year}
  {2017})}\BibitemShut {NoStop}%
\bibitem [{\citenamefont {Ma}\ \emph {et~al.}(2019)\citenamefont {Ma},
  \citenamefont {Saxberg}, \citenamefont {Owens}, \citenamefont {Leung},
  \citenamefont {Lu}, \citenamefont {Simon},\ and\ \citenamefont
  {Schuster}}]{Ma2019}%
  \BibitemOpen
  \bibfield  {author} {\bibinfo {author} {\bibfnamefont {R.}~\bibnamefont
  {Ma}}, \bibinfo {author} {\bibfnamefont {B.}~\bibnamefont {Saxberg}},
  \bibinfo {author} {\bibfnamefont {C.}~\bibnamefont {Owens}}, \bibinfo
  {author} {\bibfnamefont {N.}~\bibnamefont {Leung}}, \bibinfo {author}
  {\bibfnamefont {Y.}~\bibnamefont {Lu}}, \bibinfo {author} {\bibfnamefont
  {J.}~\bibnamefont {Simon}},\ and\ \bibinfo {author} {\bibfnamefont {D.~I.}\
  \bibnamefont {Schuster}},\ }\bibfield  {title} {\bibinfo {title} {A
  dissipatively stabilized {{Mott}} insulator of photons},\ }\href
  {https://doi.org/10.1038/s41586-019-0897-9} {\bibfield  {journal} {\bibinfo
  {journal} {Nature}\ }\textbf {\bibinfo {volume} {566}},\ \bibinfo {pages}
  {51} (\bibinfo {year} {2019})}\BibitemShut {NoStop}%
\bibitem [{\citenamefont {{Hacohen-Gourgy}}\ \emph {et~al.}(2015)\citenamefont
  {{Hacohen-Gourgy}}, \citenamefont {Ramasesh}, \citenamefont {De~Grandi},
  \citenamefont {Siddiqi},\ and\ \citenamefont
  {Girvin}}]{hacohen-gourgy_cooling_2015}%
  \BibitemOpen
  \bibfield  {author} {\bibinfo {author} {\bibfnamefont {S.}~\bibnamefont
  {{Hacohen-Gourgy}}}, \bibinfo {author} {\bibfnamefont {V.~V.}\ \bibnamefont
  {Ramasesh}}, \bibinfo {author} {\bibfnamefont {C.}~\bibnamefont {De~Grandi}},
  \bibinfo {author} {\bibfnamefont {I.}~\bibnamefont {Siddiqi}},\ and\ \bibinfo
  {author} {\bibfnamefont {S.~M.}\ \bibnamefont {Girvin}},\ }\bibfield  {title}
  {\bibinfo {title} {Cooling and {{Autonomous Feedback}} in a
  {{Bose}}-{{Hubbard Chain}} with {{Attractive Interactions}}},\ }\href
  {https://doi.org/10.1103/PhysRevLett.115.240501} {\bibfield  {journal}
  {\bibinfo  {journal} {Physical Review Letters}\ }\textbf {\bibinfo {volume}
  {115}},\ \bibinfo {pages} {240501} (\bibinfo {year} {2015})}\BibitemShut
  {NoStop}%
\bibitem [{\citenamefont {Magnard}\ \emph {et~al.}(2020)\citenamefont
  {Magnard}, \citenamefont {Storz}, \citenamefont {Kurpiers}, \citenamefont
  {Sch{\"a}r}, \citenamefont {Marxer}, \citenamefont {L{\"u}tolf},
  \citenamefont {Walter}, \citenamefont {Besse}, \citenamefont {Gabureac},
  \citenamefont {Reuer}, \citenamefont {Akin}, \citenamefont {Royer},
  \citenamefont {Blais},\ and\ \citenamefont
  {Wallraff}}]{magnard_microwave_2020}%
  \BibitemOpen
  \bibfield  {author} {\bibinfo {author} {\bibfnamefont {P.}~\bibnamefont
  {Magnard}}, \bibinfo {author} {\bibfnamefont {S.}~\bibnamefont {Storz}},
  \bibinfo {author} {\bibfnamefont {P.}~\bibnamefont {Kurpiers}}, \bibinfo
  {author} {\bibfnamefont {J.}~\bibnamefont {Sch{\"a}r}}, \bibinfo {author}
  {\bibfnamefont {F.}~\bibnamefont {Marxer}}, \bibinfo {author} {\bibfnamefont
  {J.}~\bibnamefont {L{\"u}tolf}}, \bibinfo {author} {\bibfnamefont
  {T.}~\bibnamefont {Walter}}, \bibinfo {author} {\bibfnamefont {J.-C.}\
  \bibnamefont {Besse}}, \bibinfo {author} {\bibfnamefont {M.}~\bibnamefont
  {Gabureac}}, \bibinfo {author} {\bibfnamefont {K.}~\bibnamefont {Reuer}},
  \bibinfo {author} {\bibfnamefont {A.}~\bibnamefont {Akin}}, \bibinfo {author}
  {\bibfnamefont {B.}~\bibnamefont {Royer}}, \bibinfo {author} {\bibfnamefont
  {A.}~\bibnamefont {Blais}},\ and\ \bibinfo {author} {\bibfnamefont
  {A.}~\bibnamefont {Wallraff}},\ }\bibfield  {title} {\bibinfo {title}
  {Microwave {{Quantum Link}} between {{Superconducting Circuits Housed}} in
  {{Spatially Separated Cryogenic Systems}}},\ }\href
  {https://doi.org/10.1103/PhysRevLett.125.260502} {\bibfield  {journal}
  {\bibinfo  {journal} {Physical Review Letters}\ }\textbf {\bibinfo {volume}
  {125}},\ \bibinfo {pages} {260502} (\bibinfo {year} {2020})}\BibitemShut
  {NoStop}%
\bibitem [{\citenamefont {Vermersch}\ \emph {et~al.}(2017)\citenamefont
  {Vermersch}, \citenamefont {Guimond}, \citenamefont {Pichler},\ and\
  \citenamefont {Zoller}}]{vermersch_quantum_2017}%
  \BibitemOpen
  \bibfield  {author} {\bibinfo {author} {\bibfnamefont {B.}~\bibnamefont
  {Vermersch}}, \bibinfo {author} {\bibfnamefont {P.-O.}\ \bibnamefont
  {Guimond}}, \bibinfo {author} {\bibfnamefont {H.}~\bibnamefont {Pichler}},\
  and\ \bibinfo {author} {\bibfnamefont {P.}~\bibnamefont {Zoller}},\
  }\bibfield  {title} {\bibinfo {title} {Quantum {{State Transfer}} via {{Noisy
  Photonic}} and {{Phononic Waveguides}}},\ }\href
  {https://doi.org/10.1103/PhysRevLett.118.133601} {\bibfield  {journal}
  {\bibinfo  {journal} {Physical Review Letters}\ }\textbf {\bibinfo {volume}
  {118}},\ \bibinfo {pages} {133601} (\bibinfo {year} {2017})}\BibitemShut
  {NoStop}%
\bibitem [{\citenamefont {Xiang}\ \emph {et~al.}(2017)\citenamefont {Xiang},
  \citenamefont {Zhang}, \citenamefont {Jiang},\ and\ \citenamefont
  {Rabl}}]{xiang_intracity_2017}%
  \BibitemOpen
  \bibfield  {author} {\bibinfo {author} {\bibfnamefont {Z.-L.}\ \bibnamefont
  {Xiang}}, \bibinfo {author} {\bibfnamefont {M.}~\bibnamefont {Zhang}},
  \bibinfo {author} {\bibfnamefont {L.}~\bibnamefont {Jiang}},\ and\ \bibinfo
  {author} {\bibfnamefont {P.}~\bibnamefont {Rabl}},\ }\bibfield  {title}
  {\bibinfo {title} {Intracity {{Quantum Communication}} via {{Thermal
  Microwave Networks}}},\ }\href {https://doi.org/10.1103/PhysRevX.7.011035}
  {\bibfield  {journal} {\bibinfo  {journal} {Physical Review X}\ }\textbf
  {\bibinfo {volume} {7}},\ \bibinfo {pages} {011035} (\bibinfo {year}
  {2017})}\BibitemShut {NoStop}%
\bibitem [{\citenamefont {Vogell}\ \emph {et~al.}(2017)\citenamefont {Vogell},
  \citenamefont {Vermersch}, \citenamefont {Northup}, \citenamefont {Lanyon},\
  and\ \citenamefont {Muschik}}]{vogell_deterministic_2017}%
  \BibitemOpen
  \bibfield  {author} {\bibinfo {author} {\bibfnamefont {B.}~\bibnamefont
  {Vogell}}, \bibinfo {author} {\bibfnamefont {B.}~\bibnamefont {Vermersch}},
  \bibinfo {author} {\bibfnamefont {T.~E.}\ \bibnamefont {Northup}}, \bibinfo
  {author} {\bibfnamefont {B.~P.}\ \bibnamefont {Lanyon}},\ and\ \bibinfo
  {author} {\bibfnamefont {C.~A.}\ \bibnamefont {Muschik}},\ }\bibfield
  {title} {\bibinfo {title} {Deterministic quantum state transfer between
  remote qubits in cavities},\ }\href
  {https://doi.org/10.1088/2058-9565/aa868b} {\bibfield  {journal} {\bibinfo
  {journal} {Quantum Science and Technology}\ }\textbf {\bibinfo {volume}
  {2}},\ \bibinfo {pages} {045003} (\bibinfo {year} {2017})}\BibitemShut
  {NoStop}%
\bibitem [{\citenamefont {{Gonz{\'a}lez-Tudela}}\ \emph
  {et~al.}(2015)\citenamefont {{Gonz{\'a}lez-Tudela}}, \citenamefont
  {Paulisch}, \citenamefont {Chang}, \citenamefont {Kimble},\ and\
  \citenamefont {Cirac}}]{gonzalez-tudela_deterministic_2015}%
  \BibitemOpen
  \bibfield  {author} {\bibinfo {author} {\bibfnamefont {A.}~\bibnamefont
  {{Gonz{\'a}lez-Tudela}}}, \bibinfo {author} {\bibfnamefont {V.}~\bibnamefont
  {Paulisch}}, \bibinfo {author} {\bibfnamefont {D.~E.}\ \bibnamefont {Chang}},
  \bibinfo {author} {\bibfnamefont {H.~J.}\ \bibnamefont {Kimble}},\ and\
  \bibinfo {author} {\bibfnamefont {J.~I.}\ \bibnamefont {Cirac}},\ }\bibfield
  {title} {\bibinfo {title} {Deterministic {{Generation}} of {{Arbitrary
  Photonic States Assisted}} by {{Dissipation}}},\ }\href
  {https://doi.org/10.1103/PhysRevLett.115.163603} {\bibfield  {journal}
  {\bibinfo  {journal} {Physical Review Letters}\ }\textbf {\bibinfo {volume}
  {115}},\ \bibinfo {pages} {163603} (\bibinfo {year} {2015})}\BibitemShut
  {NoStop}%
\bibitem [{\citenamefont {Albrecht}\ \emph {et~al.}(2019)\citenamefont
  {Albrecht}, \citenamefont {Henriet}, \citenamefont {{Asenjo-Garcia}},
  \citenamefont {Dieterle}, \citenamefont {Painter},\ and\ \citenamefont
  {Chang}}]{albrecht_subradiant_2019}%
  \BibitemOpen
  \bibfield  {author} {\bibinfo {author} {\bibfnamefont {A.}~\bibnamefont
  {Albrecht}}, \bibinfo {author} {\bibfnamefont {L.}~\bibnamefont {Henriet}},
  \bibinfo {author} {\bibfnamefont {A.}~\bibnamefont {{Asenjo-Garcia}}},
  \bibinfo {author} {\bibfnamefont {P.~B.}\ \bibnamefont {Dieterle}}, \bibinfo
  {author} {\bibfnamefont {O.}~\bibnamefont {Painter}},\ and\ \bibinfo {author}
  {\bibfnamefont {D.~E.}\ \bibnamefont {Chang}},\ }\bibfield  {title} {\bibinfo
  {title} {Subradiant states of quantum bits coupled to a one-dimensional
  waveguide},\ }\href {https://doi.org/10.1088/1367-2630/ab0134} {\bibfield
  {journal} {\bibinfo  {journal} {New Journal of Physics}\ }\textbf {\bibinfo
  {volume} {21}},\ \bibinfo {pages} {025003} (\bibinfo {year}
  {2019})}\BibitemShut {NoStop}%
\bibitem [{\citenamefont {Scigliuzzo}\ \emph {et~al.}(2020)\citenamefont
  {Scigliuzzo}, \citenamefont {Bengtsson}, \citenamefont {Besse}, \citenamefont
  {Wallraff}, \citenamefont {Delsing},\ and\ \citenamefont
  {Gasparinetti}}]{scigliuzzo_primary_2020-1}%
  \BibitemOpen
  \bibfield  {author} {\bibinfo {author} {\bibfnamefont {M.}~\bibnamefont
  {Scigliuzzo}}, \bibinfo {author} {\bibfnamefont {A.}~\bibnamefont
  {Bengtsson}}, \bibinfo {author} {\bibfnamefont {J.-C.}\ \bibnamefont
  {Besse}}, \bibinfo {author} {\bibfnamefont {A.}~\bibnamefont {Wallraff}},
  \bibinfo {author} {\bibfnamefont {P.}~\bibnamefont {Delsing}},\ and\ \bibinfo
  {author} {\bibfnamefont {S.}~\bibnamefont {Gasparinetti}},\ }\bibfield
  {title} {\bibinfo {title} {Primary {{Thermometry}} of {{Propagating
  Microwaves}} in the {{Quantum Regime}}},\ }\href
  {https://doi.org/10.1103/PhysRevX.10.041054} {\bibfield  {journal} {\bibinfo
  {journal} {Physical Review X}\ }\textbf {\bibinfo {volume} {10}},\ \bibinfo
  {pages} {041054} (\bibinfo {year} {2020})}\BibitemShut {NoStop}%
\bibitem [{\citenamefont {Lu}\ \emph {et~al.}(2021)\citenamefont {Lu},
  \citenamefont {Bengtsson}, \citenamefont {Burnett}, \citenamefont {Wiegand},
  \citenamefont {Suri}, \citenamefont {Krantz}, \citenamefont {Roudsari},
  \citenamefont {Kockum}, \citenamefont {Gasparinetti}, \citenamefont
  {Johansson},\ and\ \citenamefont {Delsing}}]{lu_characterizing_2021-2}%
  \BibitemOpen
  \bibfield  {author} {\bibinfo {author} {\bibfnamefont {Y.}~\bibnamefont
  {Lu}}, \bibinfo {author} {\bibfnamefont {A.}~\bibnamefont {Bengtsson}},
  \bibinfo {author} {\bibfnamefont {J.~J.}\ \bibnamefont {Burnett}}, \bibinfo
  {author} {\bibfnamefont {E.}~\bibnamefont {Wiegand}}, \bibinfo {author}
  {\bibfnamefont {B.}~\bibnamefont {Suri}}, \bibinfo {author} {\bibfnamefont
  {P.}~\bibnamefont {Krantz}}, \bibinfo {author} {\bibfnamefont {A.~F.}\
  \bibnamefont {Roudsari}}, \bibinfo {author} {\bibfnamefont {A.~F.}\
  \bibnamefont {Kockum}}, \bibinfo {author} {\bibfnamefont {S.}~\bibnamefont
  {Gasparinetti}}, \bibinfo {author} {\bibfnamefont {G.}~\bibnamefont
  {Johansson}},\ and\ \bibinfo {author} {\bibfnamefont {P.}~\bibnamefont
  {Delsing}},\ }\bibfield  {title} {\bibinfo {title} {Characterizing
  decoherence rates of a superconducting qubit by direct microwave
  scattering},\ }\href {https://doi.org/10.1038/s41534-021-00367-5} {\bibfield
  {journal} {\bibinfo  {journal} {npj Quantum Information}\ }\textbf {\bibinfo
  {volume} {7}},\ \bibinfo {pages} {1} (\bibinfo {year} {2021})}\BibitemShut
  {NoStop}%
\bibitem [{\citenamefont {Hoi}\ \emph {et~al.}(2011)\citenamefont {Hoi},
  \citenamefont {Wilson}, \citenamefont {Johansson}, \citenamefont {Palomaki},
  \citenamefont {Peropadre},\ and\ \citenamefont
  {Delsing}}]{hoi_demonstration_2011-1}%
  \BibitemOpen
  \bibfield  {author} {\bibinfo {author} {\bibfnamefont {I.-C.}\ \bibnamefont
  {Hoi}}, \bibinfo {author} {\bibfnamefont {C.~M.}\ \bibnamefont {Wilson}},
  \bibinfo {author} {\bibfnamefont {G.}~\bibnamefont {Johansson}}, \bibinfo
  {author} {\bibfnamefont {T.}~\bibnamefont {Palomaki}}, \bibinfo {author}
  {\bibfnamefont {B.}~\bibnamefont {Peropadre}},\ and\ \bibinfo {author}
  {\bibfnamefont {P.}~\bibnamefont {Delsing}},\ }\bibfield  {title} {\bibinfo
  {title} {Demonstration of a {{Single}}-{{Photon Router}} in the {{Microwave
  Regime}}},\ }\href {https://doi.org/10.1103/PhysRevLett.107.073601}
  {\bibfield  {journal} {\bibinfo  {journal} {Physical Review Letters}\
  }\textbf {\bibinfo {volume} {107}},\ \bibinfo {pages} {073601} (\bibinfo
  {year} {2011})}\BibitemShut {NoStop}%
\bibitem [{\citenamefont {Reiter}\ and\ \citenamefont
  {S{\o}rensen}(2012)}]{reiter_effective_2012-1}%
  \BibitemOpen
  \bibfield  {author} {\bibinfo {author} {\bibfnamefont {F.}~\bibnamefont
  {Reiter}}\ and\ \bibinfo {author} {\bibfnamefont {A.~S.}\ \bibnamefont
  {S{\o}rensen}},\ }\bibfield  {title} {\bibinfo {title} {Effective operator
  formalism for open quantum systems},\ }\href
  {https://doi.org/10.1103/PhysRevA.85.032111} {\bibfield  {journal} {\bibinfo
  {journal} {Physical Review A}\ }\textbf {\bibinfo {volume} {85}},\ \bibinfo
  {pages} {032111} (\bibinfo {year} {2012})}\BibitemShut {NoStop}%
\end{thebibliography}%
\onecolumngrid
\appendix
\beginsupplement

\newpage
\section{\LARGE Supplementary Material}
\subsection{Experimental Wiring}

\begin{figure*}[ht]
    \centering
    \includegraphics[width=1\linewidth]{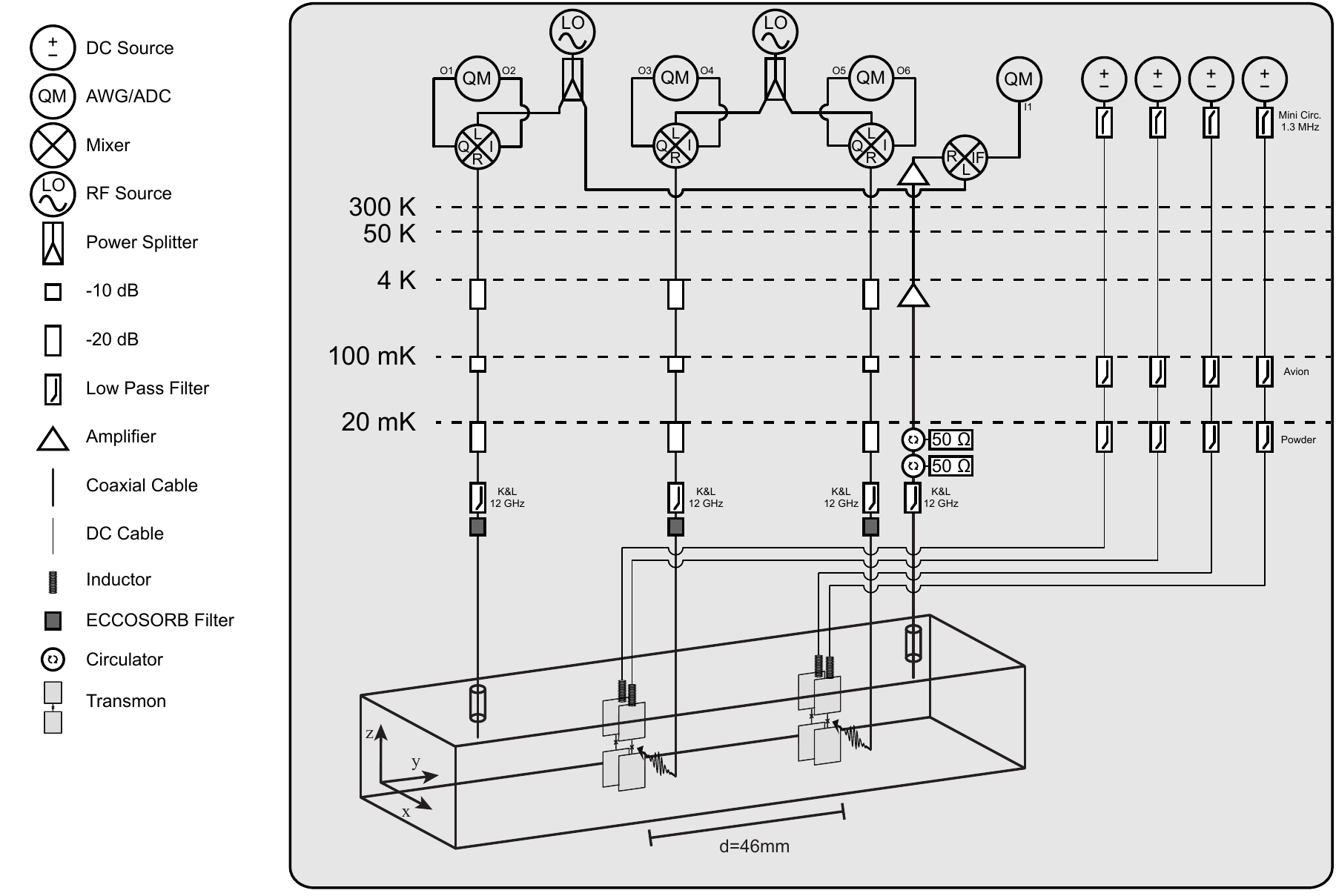}
    \caption{\textbf{Schematic of the experimental wiring with selected components.}}
    \label{fig:sup_cd_schematic}
\end{figure*}

\begin{figure*}[ht]
    \centering
    \includegraphics[width=1\linewidth]{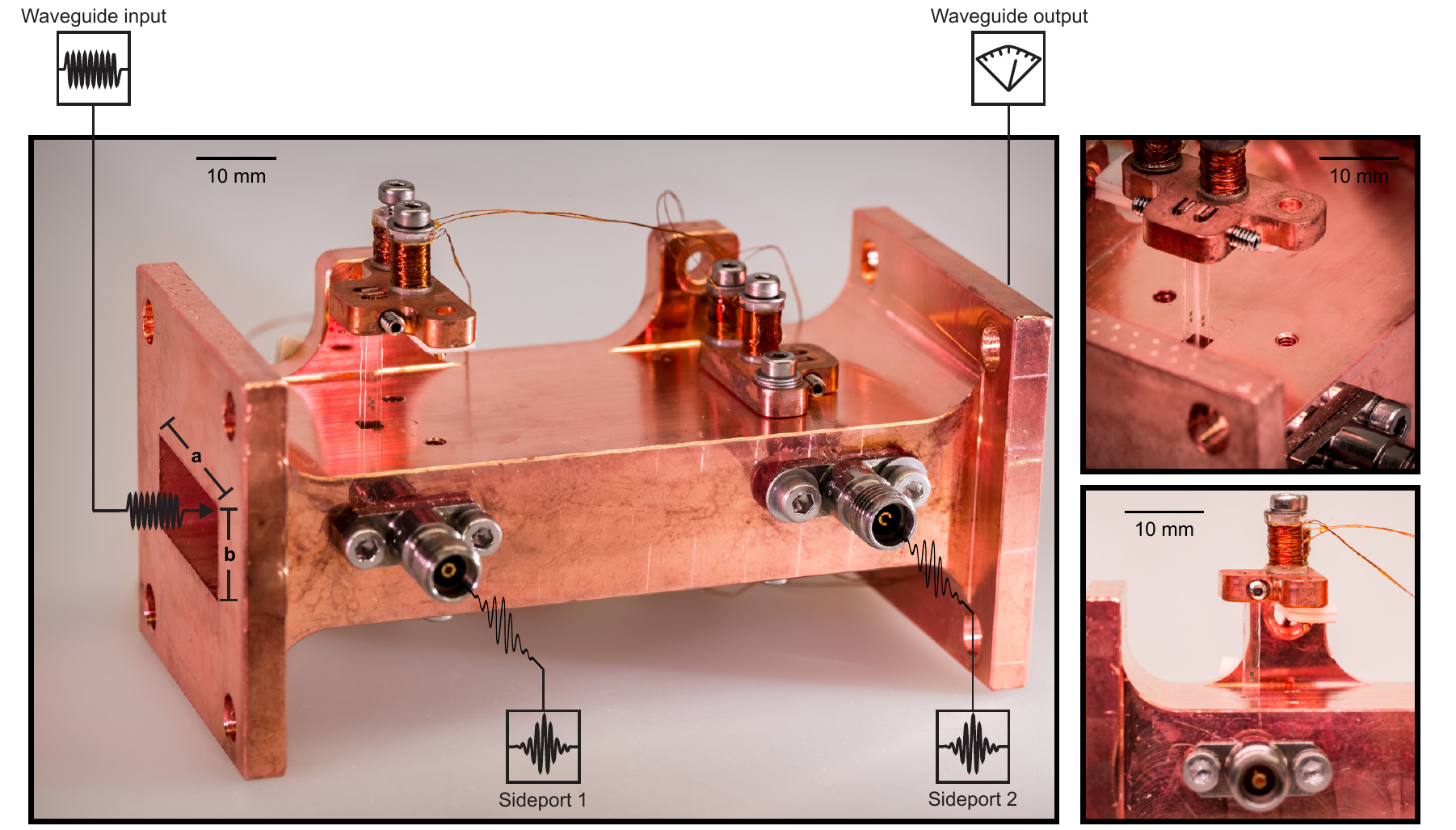}
    \caption{\textbf{Photograph of the waveguide assembly.} The waveguide middle section has one set of qubits mounted and the other aligned on the top for better visibility. The coils are mounted such that they have different couplings to the qubits. Waveguide-to-coaxial adapters are attached to the open left and right side of the waveguide, such that the transmission can be probed.}
    \label{fig:sup_waveguide}
\end{figure*}

The measurements are performed in a Triton Cryofree dilution refrigerator system with a DU7-300 dilution unit, that was able to cool to a base temperature of \SI{20}{\milli\kelvin}. Due to a malfunction of the pulsetube, the base temperature varied between \SI{25}{\milli\kelvin} and \SI{35}{\milli\kelvin} during the measurements. In Fig.~\hyperref[fig:sup_cd_schematic]{S1}, the dashed line at \SI{300}{\kelvin} separates the crysotat from the room-temperature electronics. The input coaxial cables are attenuated by \SI{20}{\decibel} at the \SI{4}{\kelvin} plate, then by \SI{10}{\decibel} at the still plate and another \SI{20}{\decibel} at the mixing chamber plate. Pulses are generated by mixing a continuous wave (CW) microwave pump ($\approx$ \SIrange{6}{8}{\giga \hertz}) with a modulated signal ($\approx$ \SIrange{0}{500}{\mega \hertz}) from an arbitrary waveform generator (AWG, Operator X - Quantum Machines) by an IQ mixer. The upmixing setup includes various filters, attenuators and switches to achieve the desired suppression of noise and the unwanted sidebands. The signal is finally filtered at the base plate by a 6L250-12000 low-pass filter from K\&L, followed by a custom built Eccosorb filter. The rectangular holes at the end of the middle section of the waveguide (see Fig.~\hyperref[fig:sup_waveguide]{S2}) are closed by commercial WR90 waveguide-to-coaxial adapters from Huber\&Suhner. The waveguide contains four superconducting quantum interference device (SQUID) operated in the transmon regime and four superconducting coils, such that the frequency of each transmon can be tuned individually. The sample is placed into a $\mu$-metal tube which sits inside a superconducting shield to protect the sample against stray magnetic fields. The output of the waveguide is attached to a K\&L filter which is connected to two Quinstar isolators giving \SI{40}{\decibel} isolation. The signal is amplified by a high electron mobility transistor (HEMT) at the \SI{4}{\kelvin} plate and further at room temperature to optimize for the required detection voltage. The signal is then downconverted to an intermediate frequency using an image-rejection mixer, filtered and finally digitized by the Operator X from Quantum Machines, which serves as the AWG for pulse generation and analog to digital converter (ADC) for signal detection. The frequencies of the transmons can be changed via superconducting coils, attached on top of the waveguide. DC currents are applied from Yokogawa GS210 current sources, where the DC bias lines are filtered by two commercial filters at room temperature and the \SI{100}{\milli\kelvin} stage and by a custom built dissipative filter at base.

\subsection{Waveguide and Transmons}
\label{samples}
The middle section of the rectangular waveguide is fabricated from oxygen-free high purity copper. The inner dimensions are chosen to be $\SI{10.2}{\milli\metre}\times\SI{22.9}{\milli\metre}\times\SI{100}{\milli\metre}$, such that the fundamental cutoff frequency $\omega_{c,10}/2\pi=\frac{1}{2a\sqrt{\mu\epsilon}}=\SI{6.546}{\giga\hertz}$ only depends on the longest extension $\rm a=\SI{22.9}{\milli\metre}$ perpendicular to the propagation direction and the vacuum permittivity $\epsilon$ and permeability $\mu$. For this mode the polarization of the electric field is parallel to the dipole antenna of the transmons. The next higher mode cutoffs are the TE$_{20}$ mode at $\omega_{c,20}/2\pi=\SI{13.091}{\giga\hertz}$ and TE$_{01}$ mode at $\omega_{c,01}/2\pi=\SI{14.696}{\giga\hertz}$. For frequencies above the cutoff, the electromagnetic field propagates through the hollow core of the waveguide with propagation constant $\beta=\sqrt{k^2-k_c^2}$, defined by the
wavevector of the propagating mode $\mathrm{k}=\omega\sqrt{\epsilon\mu}$ and the cutoff wavevector $\mathrm{k_c}=\pi/a$. The phase velocity is then $\mathrm{v_p}=\omega/\beta$. The wavelength in the waveguide is $\lambda_g=2\pi/\beta$.

The transmon qubit design is patterned by electron-beam lithography (Raith eLINE Plus \SI{30}{\kilo\volt}) on a bi-layer resist stack (bottom layer: $\SI{1}{\micro\meter}$ of MMA(8.5)MAA EL13, top layer: $\SI{0.2}{\micro\meter}$ of 950 PMMA A4). The substrate are sapphire wafers ($\diameter \SI{50.8}{\milli\meter}$), therefore we sputter a thin layer of gold on top of the PMMA to avoid charging of the sample. After lithography, the gold is etched in a solution of potassium iodide with iodine and water. After developing the resist in a isopropyl alcohol \& water (3:1) solution, two layers of aluminum (\SI{25}{\nano\meter} + \SI{30}{\nano\meter}) are evaporated with a Plassys MEB550S electron-beam evaporator. The junction barrier is formed by a controlled oxidation step (\SI{15}{\milli\bar} for \SI{1}{\minute}) before the deposition of the second layer. After liftoff, the samples are diced into individual chips and inserted into the waveguide. They are thermalized by a clamp that is attached to the waveguide housing. The transmons are tunable between \SI{6}{\giga\hertz} and \SI{8.5}{\giga\hertz}. The lower frequency sweetspot arises from the asymmetric junction design of the SQUID. The Josephson junctions have sizes of roughly $A_1 = \SI{0.18}{\micro\metre}\times\SI{1.4}{\micro\metre}$ and $A_2 = \SI{0.18}{\micro\metre}\times\SI{0.51}{\micro\metre}$, the squid loop encloses an area of $A_{\rm squid} = \SI{100}{\micro\metre}\times\SI{100}{\micro\metre}$. The antenna of the transmon is formed by two rectangular pads of size $A_{\rm pad} = \SI{400}{\micro\metre}\times\SI{500}{\micro\metre}$ (width $\times$ height) separated by a gap of $d_{\rm gap}= \SI{200}{\micro\metre}$ and connected by the wires leading to the junctions.


In Tab.~\hyperref[tab:transmonpars]{S1} we summarize important transmon parameters. We can extract linwidths for individual qubits and two-qubit and four-qubit bright states from the circle-fit routine~\cite{probst_efficient_2015-3}. $Q_1$ and $Q_3$ have a larger coupling rate $\gamma$ than $Q_2$ and $Q_4$ which is mainly caused by the orientation of the chips. The chips are facing the same direction and are symmetrically aligned around the center, which means that due to the width of the sapphire substrate of \SI{330}{\micro\meter}, the metallic structures are not perfectly symmetric around the waveguide center. The bright states of pairs $Q_1$ \& $Q_2$ and $Q_3$ \& $Q_4$ correct these imperfections as ($\gamma_{Q1Q2}=\SI{27.6}{\mega\hertz}$ and $\gamma_{Q3Q4}=\SI{27.9}{\mega\hertz}$).

From a circle fit routine we extract a non-radiative decay $\gamma_{\rm nr}$ larger than \SI{600}{\kilo\hertz} for all transmons. In contrast, time-resolved decay measurements for the dark state $\ket{D_3}$ give $1/(2\pi\times T_1) = \frac{1}{2\pi\cdot \SI{1.7}{\micro\second}}=\SI{94}{\kilo\hertz}$. This discrepancy can be explained as the linewidth $\Gamma$ that is extracted from the circle-fit is a measurement of the decoherence-rate, including all decay channels. The coupling strength is given by the ratio of the full linewidth and the depth of the resonance dip, such that we obtain $\gamma_{\rm nr}'=\Gamma-\gamma$. Even if we measure with low power, such that the Rabi frequency of the drive is much smaller than the decay rate, any noise resonant with the qubit transition will start to saturate the qubit~\cite{scigliuzzo_primary_2020-1}. Saturation will effectively change the ratio between $\Gamma$ and $\gamma$, which then also changes $\gamma_{\rm nr}'$. Thus, in this analysis $\gamma_{\rm nr}'$ is overestimated and serves as an upper bound. Furthermore, the theoretical expression for the dark state decay Eq.~\eqref{eq:t1_eq} includes pure dephasing $\gamma_{\phi}$ as this can cause imperfections of the symmetries (e.g. detuning from the decoherence-free frequency $\omega_{\pi}$) which causes a finite coupling to the waveguide. As demonstrated in Ref.\cite{lu_characterizing_2021-2}, an elaborate study of various decoherence mechanisms can be performed in waveguide quantum electrodynamics experiments. The dark states and their ability to in-situ tune the coupling to the waveguide can help to gain further insights into the intricate loss mechanisms of a superconducting qubit. 

Single transmon anharmonicities U (energy difference of the $\ket{e}-\ket{f}$ transition compared to $\ket{g}-\ket{e}$) are extracted from a two-tone spectroscopy, where the $\ket{g}-\ket{e}$ transition is saturated by a strong pump and then transmission is measured around the expected frequency of $\ket{e}-\ket{f}$, similar to the measurement in Ref.~\cite{hoi_demonstration_2011-1}.
The waveguide-couplings of the different transmons are designed to be equal but they differ by at most \SI{3}{\mega\hertz}.

Capacitive coupling strengths of qubits $Q_1$ \& $Q_2$ and $Q_3$ \& $Q_4$ are extracted from the avoided crossings, exemplary shown in Fig.~\hyperref[fig:Fig3]{3b} of the main text.

\begin{table*}[ht]
    \begin{tabular}{|c c c c c c|}
         \hline
         Transmons & $\Gamma/2\pi$ (MHz) &	$\gamma/2\pi$ (MHz) & $\gamma_{\rm nr}'/2\pi$ (MHz) &	U$/2\pi$ (MHz)	& $\rm J_{ij}/2\pi$ (MHz) \\ [0.5ex]
         \hline\hline
         $Q_1$   &	15.7 &	29.8 & 0.8 &	219	&-\\
         $Q_2$   &	13.4 &	25.4	& 0.6 &	222	&-\\
         $Q_3$   &	16.6 &	31.4	& 0.9 & 225	&-\\
         $Q_4$   &	13.7 &	26.0	& 0.7 &	206	&-\\
         $Q_1Q_2$ &	29.7 (29.1) & 55.2 (55.2) &	2.1 (1.5)	&- &	43\\
         $Q_1Q_3$ &	33.5 (32.2) &	63.6 (61.0) &	1.7 (1.7) &- &-\\
         $Q_1Q_4$ &	30.4 (29.4) &	57.6 (55.8) &	1.6 (1.6) &- &-\\	
         $Q_2Q_3$ &	30.6 (29.9) &	58.0 (56.8) &	1.6 (1.5) &- &-\\	
         $Q_2Q_4$ &	27.4 (27.1) &	52.0 (51.4) &	1.4 (1.4) &- &-\\		
         $Q_3Q_4$ &	30.2 (30.3) &	55.8 (57.4) &	2.2 (1.6)	&- &	47\\
         $Q_1Q_2Q_3Q_4$ &	60.9 (59.3) &	115.0 (112.6) &	3.4 (3.1) &- &-\\
         \hline
    
    \end{tabular}
    \caption{Single qubit parameters and circle fit results. In brackets the number corresponds to the value of the added single qubit linewidths. We use the values of $\gamma$, $U$ and $J_{ij}$ in the numerical simulations, where additionally we have to distinguish between non-radiative dissipation $\gamma_{\rm nr}/2\pi=15$~kHz, pure dephasing $\kappa_\phi/2\pi = 100$~kHz and collective dephasing $K_\phi/2\pi=437$~kHz.}
    \label{tab:transmonpars}
\end{table*}


\newpage
\subsection{Time-domain Measurements}
Relaxation and coherence times quoted in the main text $T_1 = \SI[separate-uncertainty]{1.71\pm0.06}{\micro\second}$ and $T_2 = \SI[separate-uncertainty]{0.58\pm0.06}{\micro\second}$ are the weighted average from the extracted fit-parameters of repeated measurements over a span of \SI{4}{\hour}. The error is taken from the standard deviation, where the maximal and minimal measured values are $T_{1,\rm max}=\SI{1834}{\nano\second}$, $T_{1,\rm min}=\SI{1597}{\nano\second}$, $T_{2,\rm max}=\SI{686}{\nano\second}$, $T_{2,\rm min}=\SI{462}{\nano\second}$. The characteristic times are reproducible, as we measured them again during a \SI{12}{\hour} measurement, where we recalibrated the qubit detunings and obtained similar results.

For the pulsed spectroscopy of the second excitation manifold in Fig.~\hyperref[fig:Fig4]{4} we generate microwave pulses by converting the frequency of a local oscillator (LO) to the pulse frequency via an IQ-mixer. The pulses are shaped by multiplying the LO frequency with the I and Q frequency, provided by the Operator X. The mixer creates harmonic sidebands, detuned from the LO frequency by the intermediate frequency (IF) of the I and Q signal. In order to achieve a range of \SI{400}{\mega\hertz}, we use the left sideband at frequency $f_{\rm LS} = f_{\rm LO}-f_{\rm IF}$ and the right sideband at $f_{\rm LS} = f_{\rm LO}+f_{\rm IF}$. For the measurement, pulse frequencies higher than $f_{\rm center}=\SI{7.158}{\giga\hertz}$ use the right sideband, while lower frequencies use the left sideband. To correct for the phase difference between the sidebands (\SI{180}{\degree}) we shifted the lower sideband data by $\pi$. This distorts the background and is responsible for the horizontal line at $f_{\rm center}$, separating the plot in an upper and lower half in Fig.~\hyperref[fig:Fig4]{4}.

\begin{figure*}[ht]
    \centering
    \includegraphics[width=1\linewidth]{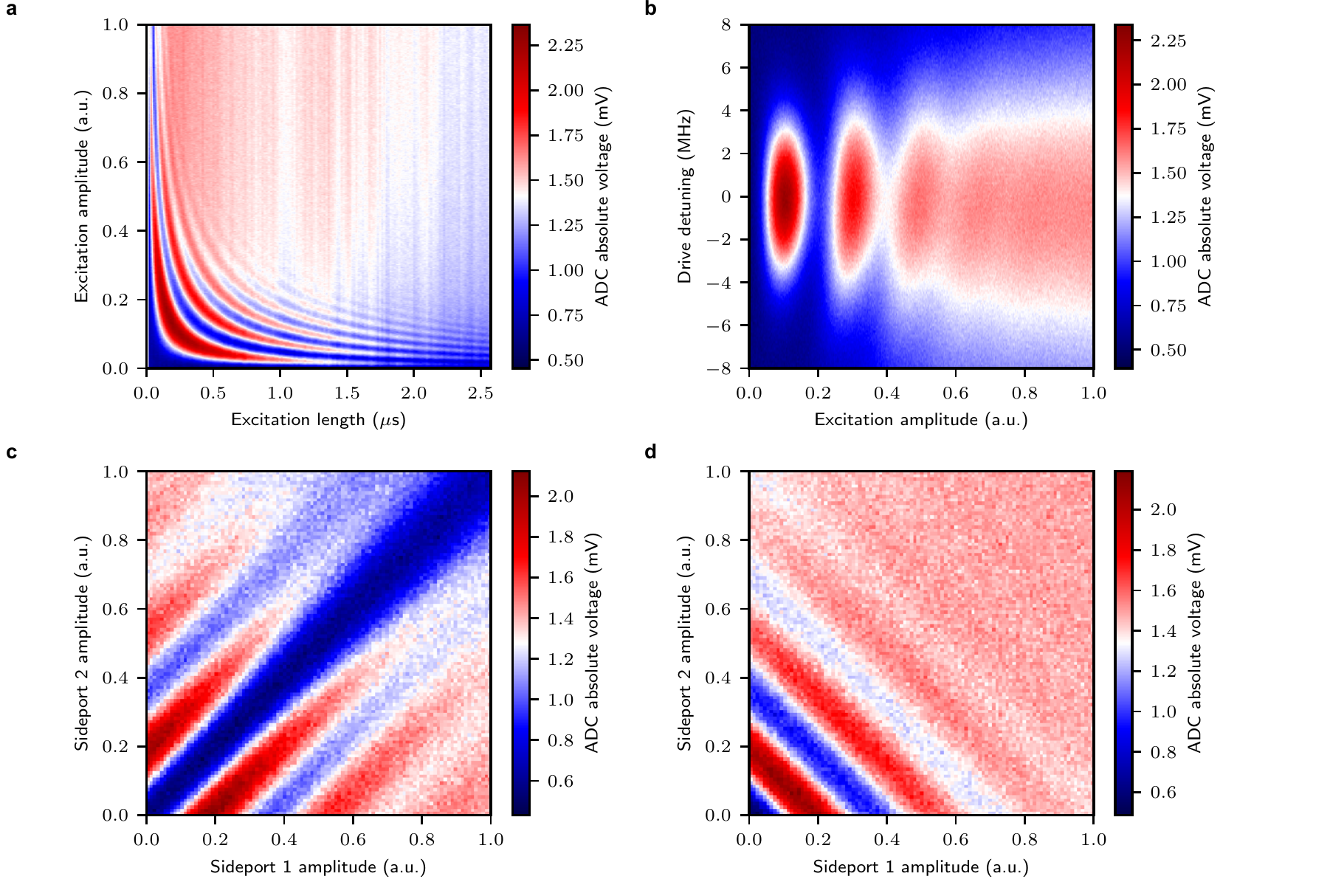}
    \caption{\textbf{Rabi-oscillations between $\ket{G}$ and $\ket{D_3}$ for different parameters.}
    \textbf{a} Varying the pulselength and amplitude we observe more oscillations for longer pulses as the effective width of the pulse in the frequency domain becomes smaller, hence a more frequency-selective driving is possible. Finite decay time T$_1$ limits to go to very long pulses. 
    \textbf{b} When detuning the drive frequency we observe an asymmetric damped oscillation, hinting other transitions at lower frequencies.
    \textbf{c} Setting the phase to $\phi= \pi$, we cannot observe Rabi-oscillations for equal pulse amplitudes on both sideports. Only when we introduce unequal drive strengths, we recover oscillations.
    \textbf{d} With the phase fixed to $\phi= 0$, we have a optimal symmetrical drive that can drive the dark state $\ket{D_3}$. As long as there is enough symmetrical part in the drive, we can drive oscillations. 
    }
    \label{fig:sup_rabis}
\end{figure*}

Fig.~\hyperref[fig:sup_rabis]{S3a} shows Rabi oscillations between the ground state $\ket{G}$ and collective dark state $\ket{D_3}$, where we increase the amplitude on the vertical axis and the length on the horizontal axis of a gaussian excitation pulse with phase between sideports $\phi=0$. The amplitude between sideports is equally increased in this measurement. Longer pulses decrease the width in frequency space and therefore lead to less driving of off-resonant transitions (mainly $\ket{3}$ to $\ket{14}$). At the same time the dark state will have decayed further back into the ground state. Fig.~\hyperref[fig:sup_rabis]{S3b} shows Rabi oscillations for a fixed pulse length of \SI{240}{\nano\second} and $\phi=0$, where we detuned the drive with respect to the transition frequency of the states $\ket{G}$ and $\ket{D_3}$. Fig.~\hyperref[fig:sup_rabis]{S3c} shows Rabi oscillations for a phase difference between the sideports 1 and 2 of $\phi=\pi$. With symmetric increase of power we cannot drive Rabi oscillations as we are mainly driving the collective bright state $\ket{B_4}$ which immediately decays back into the ground state. We can distort the drive symmetry by a power imbalance between the drive ports which shows that amplitude and phase contribute to the resulting symmetry. In Fig.~\hyperref[fig:sup_rabis]{S3d} the phase is set to $\phi=0$.

\newpage
\subsection{Theoretical Model}
\begin{figure*}[ht]
    \centering
    \includegraphics[width=1\linewidth]{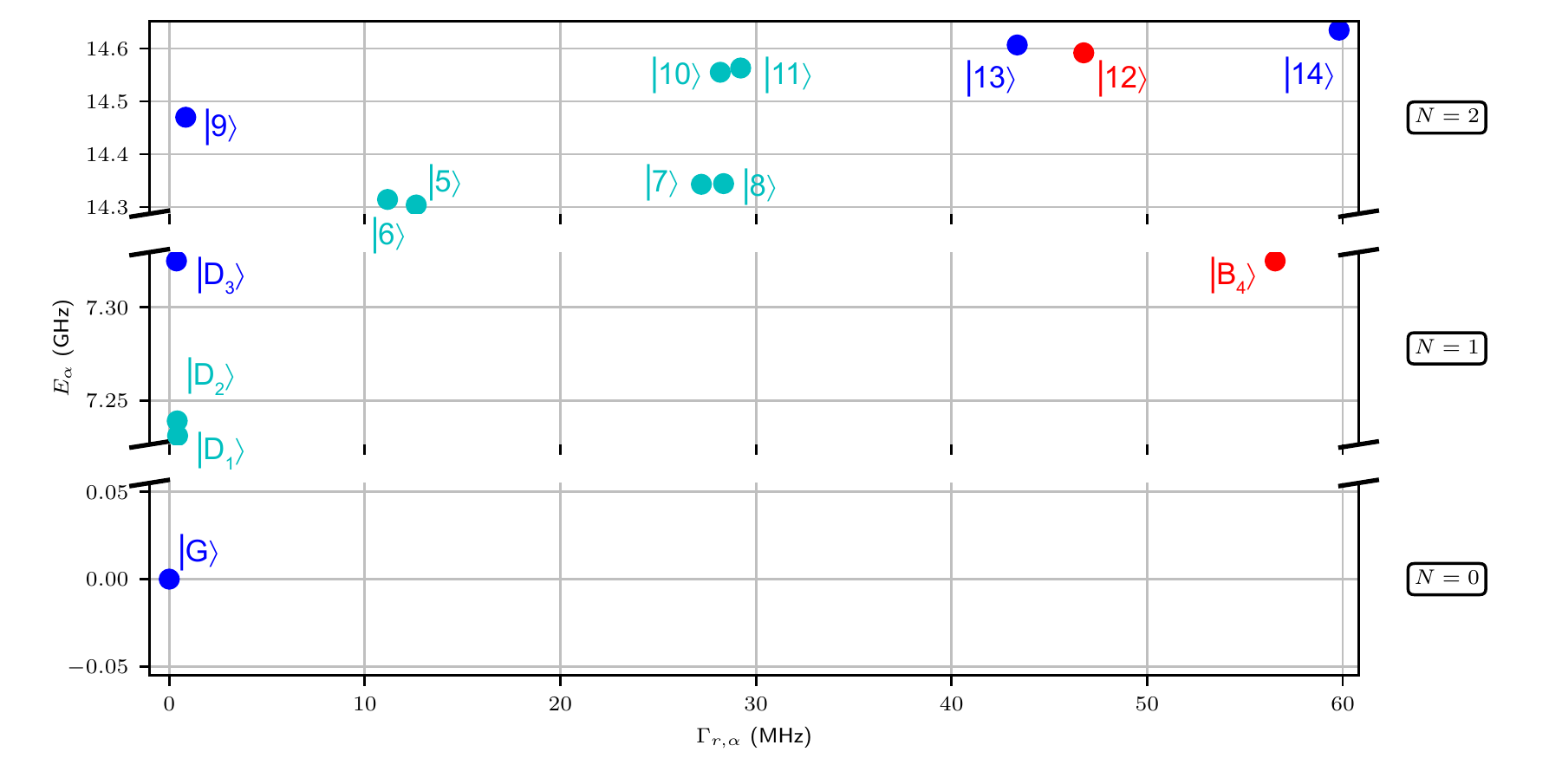}
    \caption{Numerically obtained eigenlevels in zero, one and two excitation manifolds computed with the parameters of Tab.~\ref{tab:transmonpars}. The $y$-axis gives the energy and $x$-axis the decay rate of the state. Here, states that are symmetric with respect to the exchange of pairs are colored blue and antisymmetric ones red. States with no symmetry under qubit exchange are turquoise.}
    \label{fig:sup_levels}
\end{figure*}
The dynamics of the four transmons are governed by the master equation
\begin{equation}
    \begin{aligned}
        \frac{d\hat \rho}{dt} &= -\frac{i}{\hbar}\left[
        \hat H_{\rm T} + 
        \hbar\sum_{j,k}\widetilde{J}_{j,k}\hat a_k^\dag\hat a_j,\hat \rho\right]
        +\sum_{j,k}\gamma_{j,k}\left(\hat a_j\hat\rho\hat a_k^\dag - \frac{1}{2}\hat a_k^\dag\hat a_j\hat\rho-
        \frac{1}{2}\hat\rho\hat a_k^\dag\hat a_j\right)\\
        &+\sum_j\gamma_{\rm nr}\left(\hat a_j\hat\rho\hat a_j^\dag -\frac{1}{2}\hat a_j^\dag\hat a_j\hat\rho
        -\frac{1}{2}\hat\rho\hat a_j^\dag\hat a_j\right)
        +\sum_j\kappa_\phi\left(\hat n_j\hat\rho\hat n_j -\frac{1}{2}\hat n_j^2\hat\rho
        -\frac{1}{2}\hat\rho\hat n_j^2\right)
        +K_\phi\left(\hat N\hat\rho \hat N 
        -\frac{1}{2}\hat N^2\hat\rho
        -\frac{1}{2}\hat\rho\hat N^2\right),
    \end{aligned}
\end{equation}
Here the coefficients $\widetilde J_{j,k}$ and $\gamma_{j,k}$ are the coherent exchange interaction and correlated decay between sites $j$ and $k$,
\begin{align}
    \widetilde J_{j,k} &= -i\pi g_jg_k\left(\omega_je^{i\omega_j t_{jk}} - \omega_k e^{-i\omega_kt_{jk}}\right),\\
    \gamma_{j,k} &= 2\pi g_jg_k\left(\omega_je^{i\omega_j t_{jk}} + \omega_k e^{-i\omega_kt_{jk}}\right),
\end{align}
where $t_{jk}$ is the propagation time between sites $j$ and $k$ determining the phase difference, and coupling strengths $g_j$ are connected to the individual linewidths $\gamma_j$ of transmons as $g_j = \sqrt{\frac{\gamma_j}{2\pi\omega_{j}}}$.
Parameters $\gamma_{\rm nr}$ and $\kappa_\phi$ describe the non-radiative dissipation and pure dephasing of individual transmons. We also include global dephasing $K_\phi$ arising from the flux noise affecting all transmon where we have denoted the total occupation operator $\hat N = \sum_j\hat n_j$. The properties of the system are then governed by the non-Hermitian effective Hamiltonian,
\begin{equation}
    \begin{aligned}
        \hat H_{\rm eff}/\hbar &= \hat H_{\rm T}
        +\sum_{jk}\left(\widetilde J_{j,k} -\frac{i\gamma_{j,k}}{2}\right)
        \hat a_k^\dag\hat a_j
        -\frac{i}{2}\sum_j\gamma_{\rm nr}\hat a_j^\dag\hat a_j
        -\frac{i}{2}\sum_j\kappa_\phi\hat n_j^2
        -\frac{i}{2}K_\phi\hat N^2.
    \end{aligned}
\end{equation}
The eigenvalues of the effective Hamiltonian are in general complex valued, $\lambda_\alpha = E_\alpha-i\frac{\Gamma_\alpha}{2}$, where the real part $E_\alpha$ gives the energy and imaginary part $\Gamma_\alpha$ the total decay rate of the state $\ket{\alpha}$. The effective Hamiltonian commutes with the total occupation operator, and thus the eigenvalues form manifolds with integer number of quanta. The eigenvalue gives the total decay rate for the state, but one can also calculate the decay rates to individual states, which sum up to $\Gamma_\alpha$.

In the frame rotating with the drive frequency $\omega$, the drive Hamiltonian reads
\begin{equation}
    \hat H_{\rm d} = \frac{\Omega}{2}\left[e^{i\phi}
    \left(\hat a_1+\hat a_2\right) 
    + \hat a_3+\hat a_4 + \rm{h.c.}\right],
\end{equation}
where $\Omega$ is the amplitude of the drive and $\phi$ is the phase difference between the pairs. The phase $\phi$ alters the symmetry with respect to the exchange of the pairs, but the drive is always symmetric with respect to the exchange of transmons inside the pairs. The amplitude can be converted to experimental Rabi frequency~$\Omega$. By altering the phase between the pairs one can control the symmetry of the drive and thus couple different states in the neighbouring manifolds. For small amplitudes, the drive acts as a perturbation and does not alter the eigenstates. Ideally one would have equal amplitude for all sites, but in reality there is a small amplitude gradient $\Omega_{\delta}$ within the pairs. This introduces additional driving term that is always antisymmetric with respect to the exchange of transmons within the pair,
\begin{equation}
    \hat H_{\rm d, asym} = \frac{\Omega_\delta}{4}\left[e^{i\phi}
    \left(\hat a_1-\hat a_2\right) 
    + \hat a_3-\hat a_4 + \rm{h.c.}\right].
\end{equation}
This explains why we are able to see local dark states $\ket{D_1}$ and $\ket{D_2}$ in Fig.~\ref{fig:Fig4}.

Assuming that all transmons are identical and ignoring dephasing $\kappa_\phi$ and $K_\phi$, we can solve analytically the eigenstates in zero-, one- and two-excitation manifolds. The one-excitation states are obtained from the ground state with collective operators
\begin{align}
    \hat D_1^\dag &= \frac{1}{\sqrt{2}}
    \Big(\hat a_1^\dag - \hat a_2^\dag\Big),\\
    \hat D_2^\dag &= \frac{1}{\sqrt{2}}
    \Big(\hat a_3^\dag - \hat a_4^\dag\Big),\\
    \hat D_3^\dag &= \frac{1}{2}
    \Big(\hat a_1^\dag + \hat a_2^\dag 
    + \hat a_3^\dag + \hat a_4^\dag\Big),\\
    \hat B_4^\dag &= \frac{1}{2}
    \Big(-\hat a_1^\dag - \hat a_2^\dag 
    + \hat a_3^\dag + \hat a_4^\dag\Big).
\end{align}
The states in the zero- and one-excitation manifold are then
\begin{align}
    \ket{G} = &\ket{0000} = \ket{00;00},\\
    \ket{D_1} = &\frac{1}{\sqrt{2}}
    \Big(\ket{1000}-\ket{0100}\Big) = \ket{00;10},\\
    \ket{D_2} = &\frac{1}{\sqrt{2}}
    \Big(\ket{0010}-\ket{0001}\Big) = \ket{00;01},\\
    \ket{D_3} = &\frac{1}{2}
    \Big(\ket{1000}+\ket{0100}+\ket{0010}+\ket{0001}\Big)
     = \ket{01;00},\\
    \ket{B_4} = &\frac{1}{2}
    \Big(-\ket{1000}-\ket{0100}+\ket{0010}+\ket{0001}\Big) 
    = \ket{10;00},
\end{align}
where we have used two different bases. State $\ket{n_1n_2n_3n_4}$ is the Fock basis, where $n_j$ is the number of excitations in the $j$th transmon. In state $\ket{B_4D_3;D_1D_2}$, on the other hand, $B_4$, $D_3$, $D_1$ and $D_2$ refer to the number of excitation created by the collective operators $\hat B$, $\hat D_3$, $\hat D_1$ and $\hat D_2$, respectively. In the absence of anharmonicity $U$ the eigenstates in the two-excitation manifold would be obtained by operating twice with the collective operators. Because of the anharmonicity, the real eigenstates are linear combinations of these states:
\begin{align}
    \ket{5} &= \frac{2i\gamma-4J+\sqrt{U^2+16\left(
    J-\frac{i\gamma}{2}\right)^2}}{\sqrt{2}U}\Big(
    \ket{00;20} - \ket{00;02}\Big) + \ket{11;00},\\
    \ket{6} &= c_1\Big(\ket{00;20} + \ket{00;02}\Big) + c_2\ket{02;00} + c_3\ket{20;00},\\
    \ket{7} &=
    \frac{-2i\gamma+\sqrt{U^2-4\gamma^2}}{U}\ket{01;10}
    +\ket{10;10},\\
    \ket{8} &= 
    \frac{2i\gamma-\sqrt{U^2-4\gamma^2}}{U}\ket{01;01}
    +\ket{10;01},\\
    \ket{9} &= \ket{00;11},\\
    \ket{10} &= 
    \frac{-2i\gamma-\sqrt{U^2-4\gamma^2}}{U}\ket{01;10}
    +\ket{10;10},\\
    \ket{11} &= 
    \frac{2i\gamma+\sqrt{U^2-4\gamma^2}}{U}\ket{01;01}
    +\ket{10;01},\\
    \ket{12} &= \frac{2i\gamma-4J-\sqrt{U^2+16\left(
    J-\frac{i\gamma}{2}\right)^2}}{\sqrt{2}U}\Big(
    \ket{00;20} - \ket{00;02}\Big) + \ket{11;00},\\
    \ket{13} &= b_1\Big(\ket{00;20} + \ket{00;02}\Big) + b_2\ket{02;00} + b_3\ket{20;00},\\
    \ket{14} &= a_1\Big(\ket{00;20} + \ket{00;02}\Big) + a_2\ket{02;00} + a_3\ket{20;00}.
\end{align}
Exact forms for states $\ket{6}$, $\ket{13}$ and $\ket{14}$ are omitted for simplicity, and we have also omitted normalization. Writing states $\ket{B_4D_3;D_1D_2}$ in terms of the original Fock states shows that states $\ket{20;00}$ and $\ket{02;00}$ are symmetric and $\ket{11;00}$ antisymmetric with respect to the exchange of the pairs. On the other hand, combinations of the local states $\ket{00;20}\pm\ket{00;02}$ are symmetric (antisymmetric). Combinations of local collective states, such as $\ket{10;01}$, are neither symmetric or antisymmetric. Thus, states $\ket{6}$, $\ket{9}$, $\ket{13}$ and $\ket{14}$ are symmetric and states $\ket{5}$ and $\ket{12}$ antisymmetric. Asymmetry of the transmon parameters removes the symmetry of states $\ket{5}$ and $\ket{6}$, which explains why both states are visible roughly at the same phase in Fig.~\ref{fig:Fig4}.

The lifetime $T_1$ and the coherence time $T_2$ can be measured for the dark state. There are multiple different decay processes that contribute to these. The master equation in the zero and one excitation manifolds can be solved analytically to some degree, if the transmons are identical. The correlations between the ground state and dark state evolve in time according to
\begin{equation}
    \rho_{03}(t) = \rho_{03}(0)e^{-it(\omega+J)}
    e^{-t\frac{\gamma_{\rm nr}+\kappa_\phi+K_\phi}{2}},
\end{equation}
from which we recover the coherence time as
\begin{equation}\label{eq:t2_eq}
    \frac{1}{T_2} = \frac{\gamma_{\rm nr} + \kappa_\phi + K_{\phi}}{2}.
\end{equation}
The lifetime $T_1$ is actually measured using the ground state population. The time evolution is solved from the master equation (assuming that the system is initially in the dark state):
\begin{equation}
    1-p_0(t) \approx e^{-t\left(2\gamma+\gamma_{\rm nr}+\frac{\kappa_\phi}{2}-
    \frac{1}{2}\sqrt{16\gamma^2+4\gamma\kappa_\phi + \kappa_\phi^2}\right)},
\end{equation}
from which we recover the lifetime
\begin{equation}\label{eq:t1_eq}
    \frac{1}{T_1} = 2\gamma+\gamma_{\rm nr}+\frac{\kappa_\phi}{2}
    -\frac{1}{2}\sqrt{16\gamma^2+4\gamma\kappa_\phi+\kappa_\phi^2}.
\end{equation}
Thus, we conclude that the coherence time of the dark state $\ket{D_3}$ depends on the nonradiative decay $\gamma_{\rm nr}$ as well as pure local and global dephasings $\kappa_\phi$ and $K_\phi$. Interestingly also the lifetime depends on the local dephasing. This happens because the local dephasing causes transitions from the dark state $\ket{D_3}$ to local dark states $\ket{D_1}$ and $\ket{D_2}$, as well as to the bright state $\ket{B_4}$, which decays through the waveguide.

\subsection{Optimizing the dark state manifold protection}
In Fig.~\ref{fig:Fig3}, we observe that the Rabi-drive between the ground state $\ket{G}$ and the dark state $\ket{D_3}$ excites also the states in the second excitation manifold, mainly the states $\ket{13}$ and $\ket{14}$ that subsequently decay to the bright state $\ket{B_4}$. Let us now for simplicity consider only the state $\ket{14}$ in the second excitation manifold. All the results apply also for the state $\ket{13}$. Other states are only very weakly coupled to the state $\ket{D_3}$ either by symmetry exclusion or energy difference. With the experimental parameters, the anharmonicity, i.e. the energy difference between the transition energies $\widetilde{U}/2\pi=[(E_{14}-E_{D_3})-(E_{D_3}-E_{G})]/2\pi\approx -\SI{15}{\mega\hertz}$, is of the same order as the linewidth of the state $\gamma/2\pi\approx~\SI{59.8}{\mega\hertz}$ (see Fig.~\ref{fig:sup_levels}), explaining why population can leak from the state $\ket{D_3}$ to state $\ket{14}$. 

\begin{figure*}[ht]
    \centering
    \includegraphics[width=1\linewidth]{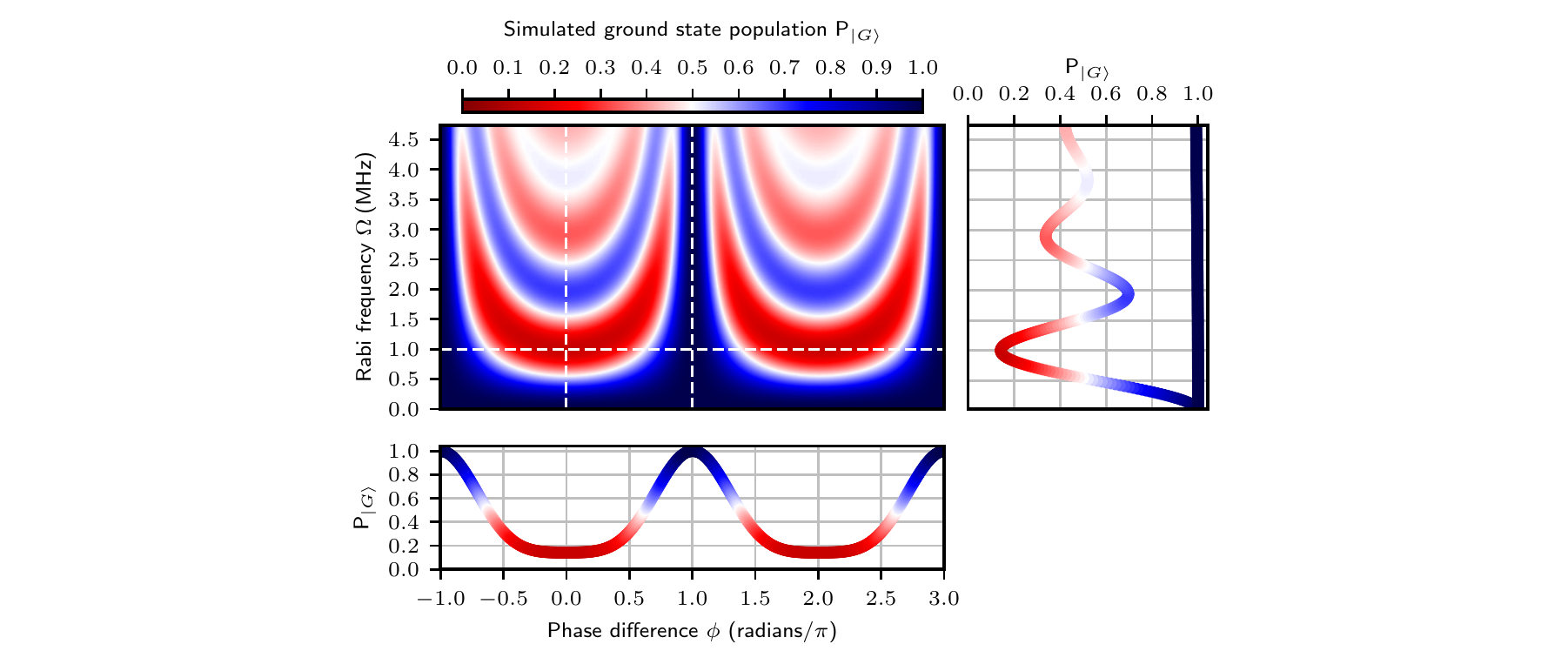}
    \caption{
    Simulated ground state population after the Rabi pulse. The dynamics of the system is solved numerically from the master equation while altering the phase and amplitude of the driving Hamiltonian. After the Rabi pulse, lasting \SI{240}{\nano\second}, the ground state population is calculated. The Rabi pulse is not perfect, since the ground state population does not go to zero.
    }
    \label{fig:sup_theory}
\end{figure*}

\begin{figure*}[ht]
    \centering
    \includegraphics[width=1\linewidth]{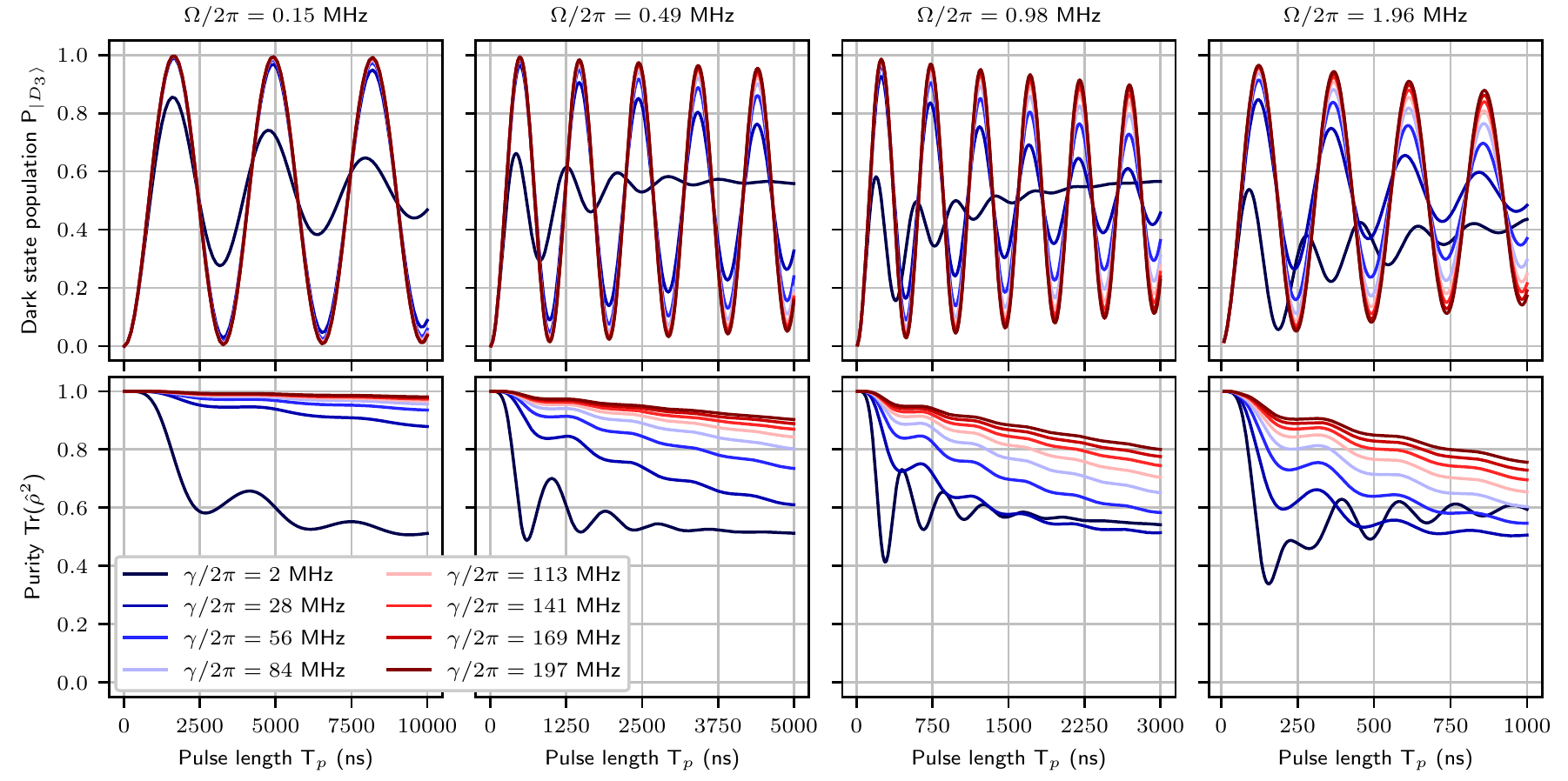}
    \caption{Simulated dark state populations (top row) and purities of the state of the system (bottom row) as a function of Rabi pulse length for different values of Rabi frequencies~$\Omega$ and waveguide couplings~$\gamma$. System parameters have been taken as average of experimental values in Tab.~\ref{tab:transmonpars}, corresponding to $U/2\pi = \SI{218}{\mega \hertz}$ and $J/2\pi = \SI{45}{\mega \hertz}$. The simulation starts from the ground state $\ket{\psi(0)}=\ket{G}$. For a perfect qubit we would expect undamped oscillations between 0 and 1 for population, and the state would remain a pure state. However, the ground state and the dark state do not form a perfect qubit because of the leakage of population to two-photon states $\ket{13}$ and $\ket{14}$, which are close to resonance with the driving frequency. There are two possibilities to improve the system and obtain better Rabi oscillations. First, one could use a weaker but longer Rabi pulse. This decreases the off-resonant driving, and more population remains in the dark state. Problem with longer pulse is non-radiative decay of the dark state, which has not been included in the simulations, but which decreases the population. Another, more surprising way is to increase the coupling to the waveguide ($\gamma/2\pi=\SI{14}{\mega\hertz}$ being the current value). We observe that the larger the~$\gamma$, the better the Rabi becomes. The decay of the population amplitude is also decreased. Similarly the state is more pure for larger~$\gamma$. However, as is evident from the results, the improvement begins to saturate, and also approximation made in the master equation would break with too large~$\gamma$. On the other hand, for weaker coupling the Rabi becomes much worse.}
    \label{fig:sup_rabi_purity}
\end{figure*}

With single solid-state qubits such as transmons~\cite{koch_charge-insensitive_2007-2}, the leakage can be minimized either by driving with a smaller Rabi drive amplitude or by engineering larger anharmonicities between the computational state and the higher excitation states. Here, we have an additional possibility: engineering the decay to the waveguide $\gamma$ so large, that the leakage is adiabatically eliminated~\cite{reiter_effective_2012-1}. In Fig.~\ref{fig:sup_rabi_purity}, we demonstrate this effect in numerical simulations of the system Hamiltonian where we use the experimental parameters of Tab.~\ref{tab:transmonpars} but assume identical qubit-waveguide coupling in the range of $\gamma/2\pi=$\SIrange{2}{197}{\mega\hertz}. An average value of $\gamma/2\pi = \SI{28}{MHz}$ thus corresponds to the experimental realization. It can be clearly seen that increasing the coupling (decay rate) to the waveguide of all the other states except the dark state, decreases leakage effects from the decoherence-free subspace resulting in a weaker damping of the Rabi oscillations between the states $\ket{G}$ and $\ket{D_3}$ and an increased overall total purity of the driven system. The coherence time $T_2$ of the dark state is independent of $\gamma$. The lifetime $T_1$ only depends weakly on $\gamma$ when it becomes large compared to the other decoherence rates, see Eqs.~\eqref{eq:t2_eq}-\eqref{eq:t1_eq}.

In the numerical simulations, we observe that the higher excitation states become only weakly excited and have negligible dynamics on the relevant timescales when the waveguide coupling is large and the Rabi frequency weak enough. In this case, the results can also be analytically explained by adiabatically eliminating the higher excitation states and reducing the dynamics only into the one excitation manifold~\cite{reiter_effective_2012-1}. Let us consider the ground state $\ket{G}$, the first excited manifold which is made of now only the bright state $\ket{B_4}$ and the dark state $\ket{D_3}$ for simplicity as well as a single state from the second excited manifold $\ket{f}$ (which can be either $\ket{13}$ or $\ket{14}$) to which the dark state is coupled through the Rabi drive. The Hamiltonian reads 
\begin{equation}
\hat H = \omega_1\ket{B_4}\bra{B_4}+\omega_1\ket{D_3}\bra{D_3}+(2\omega_1-\widetilde{U})\ket{f}\bra{f}.
\end{equation}
We drive the system with the Rabi drive that couples the ground state to the dark state, and the dark state to the state $\ket{f}$:
\begin{equation}
  \hat H_d(t)=2\Omega \cos(\omega t)\left(\ket{G}\bra{D_3}+\ket{D_3}\bra{G}\right)+2\widetilde{\Omega}\cos(\omega t)\left(\ket{D_3}\bra{f}+\ket{f}\bra{D_3}\right).
\end{equation}
We choose to drive the system at resonance $\omega=\omega_1$, resulting in the driven Hamiltonian in the rotating frame
\begin{equation}
  \hat H'=\Omega \left(\ket{G}\bra{D_3}+\ket{D_3}\bra{G}\right)+\widetilde{\Omega}\left(\ket{D_3}\bra{f}+\ket{f}\bra{D_3}\right)-\widetilde{U}\ket{f}\bra{f}.
\end{equation}
In addition to the drives we include the decay rates of the states $\ket{B_4}$ and $\ket{f}$ represented through the master equation
\begin{equation}
  \dot{\hat \rho}=-\frac{i}{\hbar}[\hat H',\hat \rho]+\left(\hat{L}^{}_B\hat\rho \hat L^\dagger_B-\frac 1 2 \hat\rho \hat L^\dagger_B\hat{L}^{}_B-\frac 1 2 \hat L^\dagger_B\hat{L}^{}_B\hat\rho \right)+\left(\hat{L}^{}_f\hat\rho \hat L^\dagger_f-\frac 1 2 \hat\rho \hat L^\dagger_f\hat{L}^{}_f-\frac 1 2 \hat L^\dagger_f\hat{L}^{}_f\hat\rho \right),
\end{equation}
where the jump operators describe the decay of the bright state $\hat{L}_{B}=\sqrt{\gamma_B}\ket{G}\bra{B_4}$ at rate $\gamma_B$ and the decay of the second excited state $\hat{L}_{f}=\sqrt{\gamma_f}\ket{B_3}\bra{f}$ at rate $\gamma_f$.

\begin{figure*}[ht]
    \centering
    \includegraphics[width=0.5\linewidth]{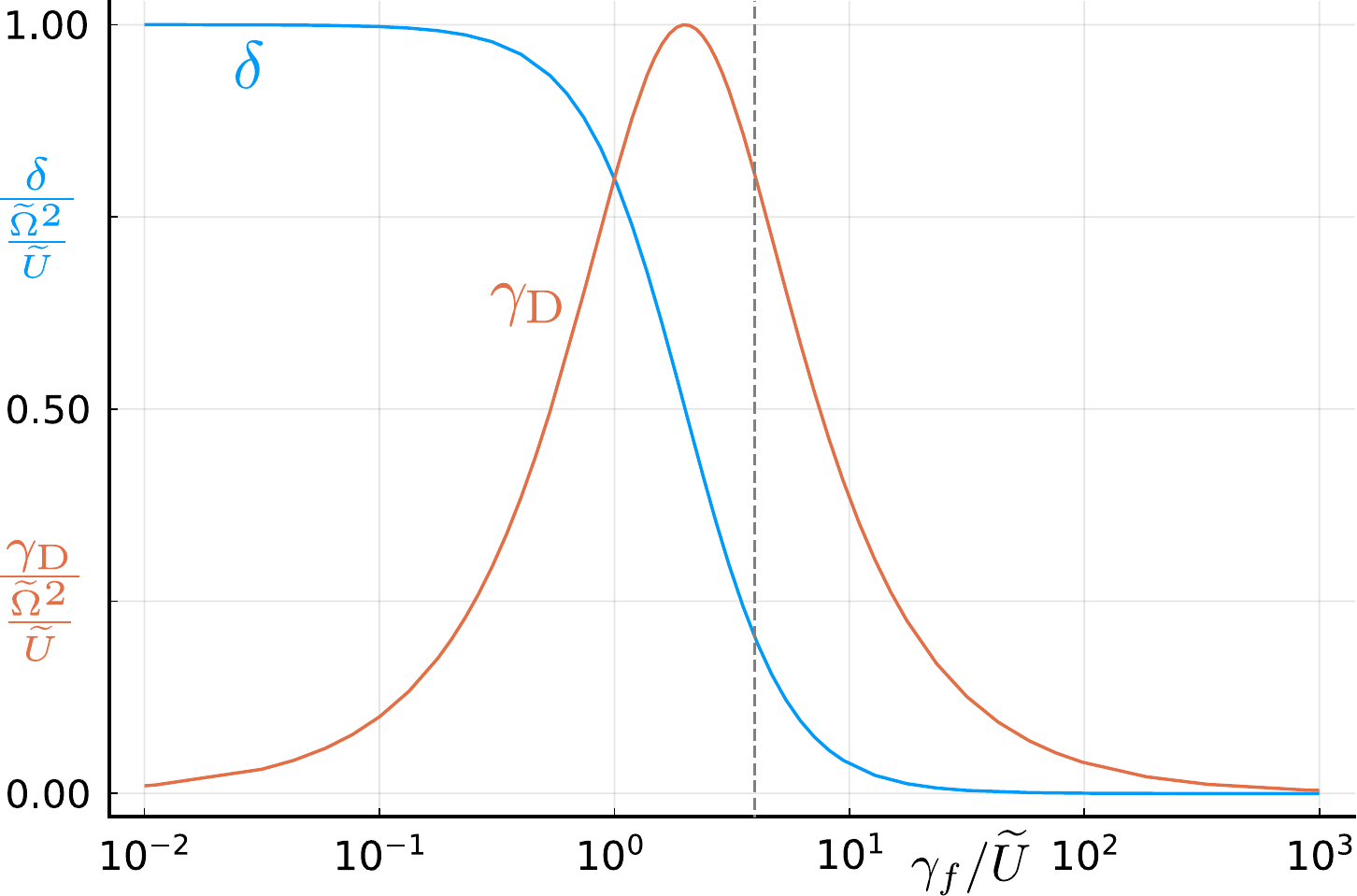}
    \caption{Effective AC Stark shift $\delta$ and the decay rate of the dark state $\gamma_{\rm D}$ as a function of the decay rate of the second excited state $\gamma_{\rm f}$ in Eq.~\eqref{eq.gammad}. The maximum of the decay rate $\gamma_{\rm D}$ occurs at $\gamma_{\rm f}/\widetilde{U}=2$. The experimental value corresponding to the state $\ket{14}$ is indicated by vertical dashed line $\gamma_{\rm f}/\widetilde{U}\approx 3.95$.}
    \label{fig:sup_elimination}
\end{figure*}

By following Ref.~\onlinecite{reiter_effective_2012-1}, we adiabatically eliminate the state $\ket{f}$ resulting in the effective Hamiltonian 
\begin{equation}
  \hat H^{\rm eff}=\Omega\left(\ket{G}\bra{D_3}+\ket{D_3}\bra{G}\right) + \frac {4\widetilde{\Omega}^2 \widetilde{U}}{4\widetilde{U}^2+\gamma^2_f}\ket{D_3}\bra{D_3},
\end{equation}
where the energy of the dark state is AC Stark shifted by $\delta=\frac {4\widetilde{\Omega}^2 \widetilde{U}}{4\widetilde{U}^2+\gamma^2_f} $. The dark state can also now decay through the bright state 
\begin{equation}
 \hat{L}_{D}=\sqrt{\gamma_f\frac{4\widetilde{\Omega}^2}{\gamma^2_f+4\widetilde{U}^2}}e^{i \theta}\ket{B_4}\bra{D_3}=\sqrt{\gamma_D}e^{i\theta}\ket{B_4}\bra{D_3}
\end{equation}
at the rate $\gamma_{\rm D}=\gamma_f\frac{4\widetilde{\Omega}^2}{\gamma^2_f+4\widetilde{U}^2}$. Notice that if the decay rate of the excited state dominates over the detuning $\gamma_f\gg \widetilde{U}$ then both the ac Stark shift and the dark state decay rate decrease as a function of the excited state decay rate $\gamma_{\rm f}$ (see Fig.~\ref{fig:sup_elimination}):
\begin{align}
  \delta=\frac {4\widetilde{\Omega}^2 \widetilde{U}}{4\widetilde{U}^2+\gamma^2_f} &=\begin{cases} \widetilde{\Omega} \frac{\widetilde{\Omega}}{\widetilde{U}}, &  \frac{\widetilde{U}}{\gamma_{\rm f}}\gg 1  \\
\widetilde{\Omega} \frac{\widetilde{\Omega}}{\widetilde{U}} \left(\frac{2\widetilde{U}}{\gamma_{f}}\right)^2, & \frac{\widetilde{U}}{\gamma_{\rm f}}\ll 1 \end{cases}, &  \gamma_D=\gamma_f\frac{4\widetilde{\Omega}^2}{\gamma^2_f+4\widetilde{U}^2}&=\begin{cases} \gamma_{f} \left(\frac{\widetilde{\Omega}}{\widetilde{U}}\right)^2, &  \frac{\widetilde{U}}{\gamma_{\rm f}}\gg 1  \\
 \gamma_{f} \left(\frac{\widetilde{\Omega}}{\widetilde{U}}\right)^2\left(\frac{2\widetilde{U}}{\gamma_{f}}\right)^2, & \frac{\widetilde{U}}{\gamma_{\rm f}}\ll 1. \end{cases} \label{eq.gammad}
\end{align}
The reduction of $\gamma_{D}$ as a function of $\gamma_{\rm f}$ (which is a function of $\gamma$) is seen in the simulations of Fig.~\ref{fig:sup_rabi_purity}.
\end{document}